\documentclass[12pt,preprintnumbers,amsmath,amssymb,nofootinbib]{revtex4-1}
\usepackage[colorlinks,
            linkcolor=blue,
            anchorcolor=blue,
            citecolor=blue]{hyperref}
\usepackage{natbib,float}
\usepackage{url}
\usepackage{exscale}
\usepackage{amsmath}
\usepackage{booktabs}
\usepackage{color}
\usepackage{graphicx}
\usepackage{exscale}
\usepackage{relsize}
\usepackage{graphicx}
\usepackage{dcolumn}
\usepackage{bm}
\usepackage{multirow}
\usepackage{CJK}
\usepackage{diagbox}
\usepackage{subfigure}
\usepackage{slashed}
\usepackage{makecell}
\usepackage{adjustbox}
\makeatletter

\newcommand{\Rmnum}[1]{\expandafter\@slowromancap\romannumeral #1@}

\newcommand{\ga}{\gamma}
\newcommand{\rx}{r_{\rm exp}}
\def\nn{\nonumber}
\def \mr {M_R}
\def \mk {m_K}
\def \gr {\Gamma_R}

\makeatother

\begin{document}

\title{Analysis on the composite  nature of the light scalar mesons $f_{0}(980)$ and $a_0(980)$}

\author{Ze-Qiang Wang$^1$}
\author{Xian-Wei Kang$^{1,2}$}\email{xwkang@bnu.edu.cn}
\author{J.~A.~Oller$^3$}\email{oller@um.es}
\author{Lu Zhang$^1$}

\affiliation{{$^{1}$ Key Laboratory of Beam Technology of Ministry of Education, College of Nuclear Science and Technology, Beijing Normal University, Beijing 100875, China\\
$^{2}$ Beijing Radiation Center, Beijing 100875, China\\
$^{3}$ Departamento de F\'{\i}sica, Universidad de Murcia, E-30071 Murcia, Spain\\
}}

\begin{abstract}

 We study the weight or compositeness of the  $\pi\pi$-$K\bar{K}$ and $\pi\eta$-$K\bar{K}$ in the composition of the $f_0(980)$ and $a_0(980)$ resonances, respectively.
 Either we use the saturation of the total width and compositeness, or we use  a Flatt\'e parametrization taking also into account the spectral function of a near-threshold resonance. We make connections and compare between these two  methods. We take input values for  the pole mass and width 
 and, in addition, for the total compositeness or the decay-width branching ratio to the lighter channel for each resonance.  It turns out that for the poles considered  the meson-meson components  are dominant for the  $f_0(980)$, while for the $a_0(980)$ resonance they  are subdominant. We also provide partial decay widths and partial compositeness coefficients, so that the $K\bar{K}$ component is the most important one for the $f_0(980)$. Additionally, this study stresses the need to distinguish between the bare and dressed couplings and widths in a Flatt\'e parametrization.  We elaborate on the connection between the partial-decay widths calculated in terms of the dressed couplings and the actual measured ones. Due to the coupled-channel dynamics when the pole lies near the heavier threshold in the second Riemann sheet some changes are needed with respect to standard relations.

\end{abstract}

\maketitle

\tableofcontents
\section{Introduction}

The nonperturbative meson--meson interactions and the related scalar-meson spectroscopy is a topic of great importance.
The nature of scalar mesons is still under debate, in spite of the efforts during several decades in the past, particularly since the discovery of the resonances $f_0(980)$ \cite{Protopopescu:1973sh} and $a_0(980)$  \cite{Ammar:1968zur}.
The  scalar mesons below 1~GeV, like the $f_{0}(500)/\sigma, K^*(800)/\kappa, f_0(980), a_0(980)$, are serious candidates to comprise a $J^{PC}=0^{++}$ nonet as required in Refs.~\cite{Jaffe:1976ig,Jaffe:1976ih,vanBeveren:1986ea,Napsuciale:1998ip,Black:2000qq,Oller:2003vf,Moussallam:2011zg}.
These resonances with vacuum quantum numbers  are crucial for
the deep understanding of spontaneous chiral symmetry breaking of quantum chromodynamics (QCD), its spectroscopy and, in general, of its nonperturbative nature \cite{Moussallam:1999aq,Oller:2006xb,Oller:2007xd,Albaladejo:2010tj,Alvarez-Ruso:2009vkn,Alvarez-Ruso:2010rqm}.
Along the decades the lightest scalar resonances have been accommodated within different models like tetraquark states \cite{Jaffe:1976ig,Jaffe:1976ih,Achasov:1980tb,Achasov:1999wv,Achasov:2020aun,Vijande:2009ac}, molecular states \cite{Weinstein:1982gc,Weinstein:1983gd,Ahmed:2020kmp,Dai:2014zta,Dai:2011bs}, dynamically generated resonances \cite{Oller:1997ti,Oller:1997ng,Oller:1998hw,Oller:1998zr,Janssen:1994wn,Lohse:1990ew}, unitarized  quark models \cite{Bramon:1980ni,vanBeveren:1986ea,Tornqvist:1995kr}, linear sigma models \cite{Black:1998wt,Scadron:1997ut,Napsuciale:2004xa,Napsuciale:2004au}, etc.

For example, in Ref.~\cite{Dai:2012kf} the $a_0(980)$ is understood as a Breit-Wigner resonance, not as a dynamically generated resonance, while the $f_0(980)$ is considered as a $K\bar{K}$ bound state.
We also notice that in Ref.~\cite{Sekihara:2014qxa} the compositeness is analyzed via the $f_{0}(980)-a_{0}(980)$ mixing intensity, and it is found that the $f_{0}(980)$ and $a_{0}(980)$ cannot be simultaneously $K\bar{K}$ bound states.
The masses of the two resonances are very close and the $f_{0}(980)-a_{0}(980)$ mixing could occur via the hadronic $K\bar{K}$ loop \cite{Achasov:1979xc,Kudryavtsev:2002uu,Hanhart:2007bd,Oller:1999ag}.
Recently, there is also interest in assessing the nature of the scalar mesons by studying semileptonic decays \cite{Cheng:2017pcq,Kang:2018jzg,Kang:2013jaa}.  For some reviews, see Refs.~\cite{Close:2002zu,Klempt:2007cp,Yao:2020bxx,Zyla:2020zbs}.

In fact, a meson has typically several components \cite{Cohen:2014vta}, such as the superposition of $q\bar{q}$ and tetraquarks $qq\bar{q}\bar{q}$ \cite{tHooft:2008rus}, gluonium \cite{Narison:2000dh}, meson-meson components, etc.
The compositeness, usually denoted  by $X$, refers to the weight in the resonance state composition of the meson-meson components in the continuum part of the free spectrum \cite{Weinberg:1962hj,Oller:2017alp} that are {\it explicitly} taken into account, e.g. as channels participating in the associated  coupled-channel meson-meson scattering.
Therefore, it is  a fundamental concept that is required for a quantitative analysis on the nature of the resonance.
In contrast, the elementariness, typically called $Z$, is the weight of the bare (compact/short-range) degrees of freedom in the resonance constitution (like four quarks or gluonium), that would also include closed-channel meson-meson components not taken into account as explicit degrees of freedom, and so that $1=Z+X$.

For a bound state case, the compositeness is a positive real number \cite{Weinberg:1962hj} between 0 and 1 (as it should be), but its straightforward extension to the resonance case gives rise to  complex-valued results \cite{Oller:2017alp}.
Several extensions have been proposed \cite{Baru:2003qq,Hyodo:2011qc,Aceti:2012dd,Aceti:2014ala,Sekihara:2014kya,Matuschek:2020gqe,Albaladejo:2022sux} to end with real sensible values for the compositeness.   In this work, we use the results of Refs.~\cite{Guo:2015daa,Oller:2017alp} that  allow a probabilistic interpretation of the compositeness relation of the resonance into open channels. Studies along these lines have also been extensively done for the case of heavy-quark resonances \cite{Meissner:2015mza,Kang:2016ezb,Kang:2016jxw,Gao:2018jhk,Guo:2020vmu,Guo:2020pvt,Du:2021bgb}. In addition, we also employ the formalism based on the evaluation of the spectral density function of the bare state associated to the resonance \cite{Bogdanova:1991zz}  by using a Flatt\'e parametrization \cite{Baru:2003qq}. We then compare between this formalism and the one previously referred for the evaluation of the compositeness and elementariness, finding compatible results between them.

The $f_{0}(980)$ and $a_0(980)$ resonances couple mainly to the channels $\pi\pi$-$K\bar{K}$ and $\pi\eta$-$K\bar{K}$, respectively. 
Hence, they are also the channels that  dominate in the study for the compositeness of the resonances.
To proceed with this study, 
the main equations stem from considering the saturation of the compositeness relationship and the total width of the resonance, from which we calculate the couplings, partial compositeness coefficients and partial-decay widths.
The implication of the branching ratio to the lighter channel, which we call $r_{\rm exp}$, 
together with the reproduction of the total width,  is also explored within our compositeness formalism. This setup allows us to obtain more definite predictions for $X$, and the smaller the branching ratio the larger the resulting $X$ by  a linear relation.
In particular, for the $f_0(980)$ the branching ratio $r_{\exp}=0.52\pm 0.12$ \cite{Aubert:2006nu}, the most recent one collected in the Particle Data Group (PDG) \cite{Zyla:2020zbs} from $B$ decays to $K\pi\pi$, implies the largest $X$ ranging around  0.6--0.9 within errors.
In turn for the $a_0(980)$ the branching ratios reported recently \cite{CrystalBarrel:2019zqh,Zyla:2020zbs}, 
which also include the PDG average, 
are much larger and then $X$ calculated here is significantly smaller, around $0.2-0.4$, taking into account errors and variations in the method of calculation. This indicates that other components in addition to the meson-meson ones play an important role in the constitution of the $a_0(980)$.\footnote{However, 
it is worth keeping in mind  that  the most sophisticated theoretical studies on $\pi\eta$ scattering matched with lattice QCD \cite{Guo:2016zep,Dudek:2016cru} obtain that the $a_0(980)$ is a pole lying in a hidden Riemann sheet from the physical energy axis.
This was also obtained before in Refs.~\cite{Guo:2012yt,Guo:2012ym}.
At this point there is a caveat, because the methods used  here, or in Ref.~\cite{Baru:2003qq}, to clarify the nature of the $a_0(980)$ cannot be applied to such scenario (in which the resonance effect manifests as a strong cusp).
}
We find that for the $f_0(980)$  the $K\bar{K}$ component has a much larger partial compositeness coefficient than  the $\pi\pi$ channel. For the $a_0(980)$ it is obtained that still the $K\bar{K}$ compositeness coefficient is also larger  than the one of the $\pi\eta$,  but not overwhelmingly dominant.
These results are a verification of those already obtained in Refs.~\cite{Janssen:1994wn,Oller:1997ti}, such that  if the $\pi\eta$ channel were removed the $a_0(980)$ would disappear, while the $f_0(980)$ would keep appearing as a $K\bar{K}$ bound state.

In connection with our use of a Flatt\'e parametrization we stress the importance of distinguishing between bare and renormalized couplings and widths. The former ones are those appearing directly in the Flatt\'e parametrization, while the latter ones are associated to the actual residues of the partial-wave amplitude of interest at the pole position in the complex energy-plane.   We also show that for the present two-channel coupled scattering, when the pole lies in the second Riemann sheet, one has to modify the interpretation of the theoretically calculated partial-decay width to the lighter channel in terms of renormalized couplings (residues), and give the proper interpretation.
These two effects explain why bare partial-decay widths, often found in the literature, are much bigger than those actually measured.

For the rest of the paper, Sec.~\ref{sec.211010.1} is dedicated to elaborate
the formalism based on the saturation of the total compositeness and decay width of the resonance. In turn, Sec.~\ref{sec.201009.3} develops the method based on the use of a Flatt\'e parametrization and introduces the spectral density function for a near-threshold resonance.
Then, we apply these methods to the study of the $f_0(980)$ and $a_0(980)$ resonances, either by taking $X$ as input in Sec.~\ref{sec.201008.1}, or by using $\rx$ in Sec.~\ref{sec.211126.1}. In terms of them we typically provide the resulting partial compositeness coefficients and partial-decay widths.
Finally, concluding remarks are given in Sec.~\ref{sec.211106.1}.

\section{Formulation of the compositeness-relation and decay-width method}
\label{sec.211010.1}

For definiteness, we proceed with the discussion on the components in the nature of the $f_{0}(980)$, and develop a method to investigate its partial-decay widths, couplings and compositeness. Later on we  also apply this method to the related isovector scalar resonance $a_0(980)$.

In what follows, we consider  two main decay channels ($\pi\pi$ and $K\bar{K}$)  of the $f_0(980)$.
We follow the standard convention  such that compositeness and elementariness coefficients are written as $X$ and $Z$, respectively, with $X+Z=1$. For the case of a bound state the coefficient $Z$ corresponds to the field renormalization constant \cite{Weinberg:1962hj,Salam:1962ap}, being real and positive and less than 1 (as  $X$ is too).
The straightforward generalization for resonances of the compositeness and elementariness gives rise to complex numbers.
As mentioned in the Introduction, several variants for the compositeness of a resonance have been discussed in the literature.
Here we will follow Ref.~\cite{Guo:2015daa}, which formulates a probabilistic interpretation of the compositeness relation involving only positive and real coefficients for the resonance.
As explained in Ref.~\cite{Oller:2017alp} the compositeness $X$ arises by evaluating the expected value of the number of mesons in the resonance divided by 2 (because we are considering two-body meson states).
After the proper unitary phase transformation of the $S$-matrix, it gives the partial compositeness coefficient for  the resonance in the form \cite{Guo:2015daa}
\begin{equation}\label{equ1}
X_{i}=\Big{|}\gamma_{i}^{2}\Big{|}\Big{|}\frac{\partial G_{i}(s)}{\partial s}\Big{|}_{s=s_{R}}\,,
\end{equation}
and the subscript $i$, with $i=1$ and 2, corresponds to the $S$-wave isoscalar $\pi\pi$ and $K\bar{K}$ channels, respectively.
 The pole position in the Mandelstam variable $s$ is called $s_R$, 
\begin{equation}\label{equ21}
s_{R}=(m_{R}-\frac{i}{2}\Gamma_{R})^2\,,
\end{equation}
with $m_R$ and $\Gamma_R$ the mass and width of the resonance, respectively.
 Furthermore,  $\gamma_{i}$ is the coupling of the resonance to the channel $i$  that is extracted from the residues of the $T$ matrix at the pole position $s_R$,
\begin{equation}
\gamma^{2}_{i}=-\lim\limits_{s\to s_{R}}(s-s_{R})T(s)_{ii}\,.
\end{equation}
$G_{i}(s)$ is the unitary two-point scalar loop function for the $i_{\rm th}$ channel and it can be written in the  form \cite{Guo:2012yt}
\begin{equation}\label{equ2}
\begin{aligned}
G_{i}(s)&=\frac{1}{16\pi^{2}}\big\{a_{i}(\mu)+\log\frac{m_{2}^{2}}{\mu^{2}}-\frac{\Delta+s}{s}\log\frac{m_{2}}{m_{1}}+\frac{p_{i}}{\sqrt{s}}[\log(s-\Delta+2\sqrt{s}p_{i})\\
&+\log(s+\Delta+2\sqrt{s}p_{i})-\log(-s+\Delta+2\sqrt{s}p_{i})-\log(-s-\Delta+2\sqrt{s}p_{i})]\big\}~.\\
\end{aligned}
\end{equation}
Here $\triangle=m_{1}^{2}-m_{2}^{2}$ and $m_{1}$, $m_{2}$ are the masses of the two particles in the channel $i$. We do not take into account the isospin breaking effects and use an average mass of the charged and neutral pions, and proceed analogously for kaons too. However, such effects are expected to be negligible in our exploration. The term $a_{i}(\mu)+\log\frac{m_{2}^{2}}{\mu^{2}}$ in Eq.~\eqref{equ2} is independent of $s$ and it disappears when taking the derivative of $G_{i}(s)$ in Eq.~\eqref{equ1}. Finally in Eq.~\eqref{equ2}, $p_{i}$ is the momentum of the channel $i$,
\begin{equation}\label{eq:momentum}
p_{i}(s)=\frac{\sqrt{[s-(m_{1}+m_{2})^{2}][s-(m_{1}-m_{2})^{2}]}}{2\sqrt{s}}
\end{equation}

The total compositeness coefficient, $X = \sum\limits_{i = 1}^{n} X_i$, is the sum over the partial compositeness coefficients $X_i$,  and it must satisfy the condition $X\leq 1 $.
As discussed in more detail in Ref.~\cite{Guo:2015daa}, Eq.~\eqref{equ1}  is properly applied to the calculation of $X_i$ for the channel $i$ under the condition that the resonance 
pole 
lies in an unphysical Riemann sheet (RS) that is connected with the physical RS along an interval of the real $s$-axis (where  $s$ is the total energy squared in the center of mass reference frame), lying above the threshold for the channel $i$.\footnote{We 
  advance that for the modern and relevant determinations of the pole structures for the $f_0(980)$ and $a_0(980)$ resonances here considered there is only a pole associated to each resonance. 
}

Equation~\eqref{equ1} is very similar to that for a bound state case, see e.g.~\cite{Oller:2017alp},
\begin{equation}
X_{i}=-\gamma_{i}^{2}\frac{\partial G_{i}(s)}{\partial s}\Big{|}_{s=s_{R}}\,,
\end{equation}
with the difference  concerning the introduction of the absolute values.

It is necessary  to distinguish the RS in which $s_{R}$ lies. For the different signs of the imaginary part of $p_{1}$ and $p_{2}$ in the complex $s$-plane, we can define the four different RSs as
\begin{equation}
\begin{aligned}\label{sheets}
\text{Sheet} ~\text{\Rmnum{1}}: ~~~\text{Im}p_{1}>0, ~\text{Im}p_{2}>0
\\
\text{Sheet} ~\text{\Rmnum{2}}: ~~~\text{Im}p_{1}<0, ~\text{Im}p_{2}>0
\\
\text{Sheet} ~\text{\Rmnum{3}}: ~~~\text{Im}p_{1}<0, ~\text{Im}p_{2}<0
\\
\text{Sheet} ~\text{\Rmnum{4}}: ~~~\text{Im}p_{1}>0, ~\text{Im}p_{2}<0
\end{aligned}
\end{equation}
The RSs  $\text{\Rmnum{2}}$ and $\text{\Rmnum{3}}$ are connected  to the physical RS $\text{\Rmnum{1}}$ from the $\pi\pi$ threshold onwards up to 
and above the $K\bar{K}$ threshold in the real $s$-axis, respectively.
The threshold of the $\pi\pi$ channel is distant from the resonance mass, while the resonance location is remarkably close to the $K\bar{K}$ threshold, cf. Eq.~\eqref{equ22} below.

Next, let us discuss how
to make the analytical extrapolation from the RS \Rmnum{1} to the RSs \Rmnum{2}, \Rmnum{3}, \Rmnum{4} in order to calculate the partial compositeness coefficient $X_i$, attending to the RS in which the pole lies.
We have to cross the cut of $G_{i}(s)$ and use its continuity property  for real values of $s$ with $s> (m_{i,1} + m_{i,2})^2$, where $m_{i,1}$ and $m_{i,2}$ are the masses of the first and second particles in the $i_{\rm th}$ channel, respectively. Then, one has that  \cite{Oller:1997ti}
\begin{equation}\label{equ20}
\begin{aligned}
G_{i}^{\text{\Rmnum{2}}}(s+i\epsilon)&=G_{i}^{\text{\Rmnum{1}}}(s-i\epsilon)=G_{i}^{\text{\Rmnum{1}}}(s+i\epsilon)-2i\text{Im}G_{i}^{\text{\Rmnum{1}}}(s+i\epsilon)\\
&=G_{i}^{\text{\Rmnum{1}}}(s+i\epsilon)
+\frac{i}{8\pi}\sqrt{\frac{[s+i\epsilon-(m_{1}+m_{2})^{2}][s+i\epsilon-(m_{1}-m_{2})^{2}]}{(s+i\epsilon)^2}}\,,
\end{aligned}
\end{equation}
where the square root is calculated in the first Riemann sheet, with the argument of the radicand between 0 and $2\pi$.
 The Eq.~\eqref{equ20} can be extrapolated to any other complex value of $s$.
Thus, the RS \text{\Rmnum{1}} is obtained with $G_{1}^{\text{\Rmnum{1}}}(s), G_{2}^{\text{\Rmnum{1}}}(s)$;
the RS \Rmnum{2} corresponds to take $G_{1}^{\text{\Rmnum{2}}}(s)$, $G_{2}^{\text{\Rmnum{1}}}(s)$;
the RS \Rmnum{3} is obtained with $G_{1}^{\text{\Rmnum{2}}}(s)$, $G_{2}^{\text{\Rmnum{2}}}(s)$;
and the RS \Rmnum{4} implies $G_{1}^{\text{\Rmnum{1}}}(s)$, $G_{2}^{\text{\Rmnum{2}}}(s)$.

The crucial inputs in the evaluation of the coefficients $X_i$, Eq.~\eqref{equ1}, are the pole position and the coupling $|\gamma_i|$.
Regarding  the pole parameters of the $f_{0}(980)$, we preferentially consider the results obtained by the dispersive analysis of Ref.~\cite{GarciaMartin:2011jx} based on the use of a set of Roy-like equations called the GKPY equations \cite{Garcia-Martin:2011iqs}.
The mass and width  of the resonance calculated in Ref.~\cite{GarciaMartin:2011jx} are
\begin{equation}\label{equ22}
m_{R}=996\pm7\ \text{MeV}, ~~\Gamma_{R}= 50^{+20}_{-12}\ \text{MeV},
\end{equation}
which provides a rather accurate determination  for the mass, while the width is affected by rather large errors.
When using this pole we consider the RS \Rmnum{2} because it was found to be there in the original publication \cite{GarciaMartin:2011jx}.

In addition we consider the $f_0(980)$ pole position from Ref.~\cite{Guo:2012yt}.
This reference performs an exhaustive study of $S$- and $P$-wave meson-meson scattering by unitarizing one-loop amplitudes in $U(3)\otimes U(3)$ chiral perturbation theory \cite{DiVecchia:1980yfw,Rosenzweig:1979ay,Witten:1980sp,Kawarabayashi:1980dp,Kawarabayashi:1980uh,Kaiser:2000gs,Herrera-Siklody:1996tqr} with explicit exchange of resonances. A large amount of experimental data on different reactions is reproduced and, at the same time, the consistency of the approach is checked by properly reproducing QCD constraints from spectral sum rules and semilocal duality  as a function of the number of colors of QCD. The resulting pole of the $f_0(980)$, found also in the RS \Rmnum{2}, is
\begin{align}
\label{211107.1}
  m_{R}=978^{+7}_{-11}\ \text{MeV}, ~~\Gamma_{R}= 58^{+18}_{-22}\ \text{MeV}\,.
\end{align}
Interestingly for this case the mass of the resonance lies clearly below the $K\bar{K}$ threshold, while $m_R$ from Ref.~\cite{GarciaMartin:2011jx} in Eq.~\eqref{equ22} is above. In this way we can now explore what is the effect of such a relative arrangement of the resonance mass with respect to the two-kaon threshold.
Instead, the width of the $f_0(980)$ is rather similar in both cases.

In our considerations, we ignore the multiparticle $4\pi$ channel whose contributions are very small up to 1~GeV as obtained in phenomenological studies where it is considered \cite{Albaladejo:2008qa,Garcia-Martin:2011iqs}, or estimated theoretically in studies based on unitarizing chiral perturbation theory \cite{Salas-Bernardez:2020hua}. We simply notice as well that the electromagnetically driven two-photon decay channel has been ignored in our calculations. References~\cite{Zyla:2020zbs,Dai:2014zta,Oller:2007sh,Oller:2008kf} obtained that $\Gamma_{f_{0}(980)\rightarrow\gamma\gamma}=0.32\pm0.05~\text{MeV}$, which contributes a tiny portion of the total width for the $f_0(980)$, and
should be much smaller than the one for $\pi\pi$, and $K\bar{K}$.

Then, we sensibly assume that the total compositeness coefficient of $f_{0}(980)$ can be expressed as the sum of the $S$-wave isoscalar $\pi\pi$ and $K\bar K$ channels,
\begin{equation}
\label{equ:square2}
X=X_{1}+X_{2}=|\gamma_{1}|^{2}\Big{|} \frac{\partial G_{1}(s)}{\partial s}\Big{|}_{s=s_{R}}+|\gamma_{2}|^{2}\Big{|} \frac{\partial G_{2}(s)}{\partial s}\Big{|}_{s=s_{R}}º\,.
\end{equation}
In addition to Eq.~(\ref{equ:square2}), another main equation stems from imposing the saturation of the width of the $f_{0}(980)$. As the threshold of the $\pi\pi$ channel is distant from the resonance we use the standard formula for the partial-decay width of the $f_0(980)$ to $\pi\pi$,
\begin{equation}\label{equ33}
\Gamma_{1}=\frac{|\gamma_{1}|^{2}p_{1}(m_{R}^{2})}{8\pi m^{2}_{R}}\,,
\end{equation}
where $p_i$ is the momentum in the rest frame of the resonance, cf.~\eqref{eq:momentum} with $s=m_{R}^{2}$.

However, the $K\bar{K}$ threshold is very close to the resonance mass and the effect of the finite width of the $f_{0}(980)$ (around 50 \text{MeV}) in the $K\bar{K}$ phase space is not negligible.
Notice that even the lower limit of the $f_0(980)$ mass within its uncertainty region in Eq.~\eqref{equ22} is indeed smaller than the $K\bar{K}$ threshold. However, since the uncertainty in the mass is much smaller than the width of the resonance, this fact is easily overturn by the mass distribution of the resonance and it does not prevent the actual decay of the $f_0(980)$ to $K\bar{K}$, even when the resonance mass is below the $K\bar{K}$ threshold.

In these regards, we consider a Lorentzian mass distribution for the resonance, and the partial-decay width is written as
\begin{equation}\label{equ:square}
\Gamma_{2}=\frac{|\gamma_{2}|^{2}}{16\pi^{2}} \int_{m_{1}+m_{2}}^{+ \infty}dW\frac{p_2(W^{2})}{W^{2}}\frac{\Gamma_{R}}{(m_{R}-W)^{2}+\Gamma_{R}^2/4}
\end{equation}
In the limit $\Gamma_{R}\rightarrow0$, Eq.~(\ref{equ:square}) becomes the standard formula for the decay width.
In a practical calculation, the upper limit of integration ($+\infty$) is replaced by $m_{R}+n\Gamma_{R}$.
For example, in Ref.~\cite{Kang:2016ezb}, the value of $n = 8$ is chosen for the $Z_{b}(10610)/Z_{b}(10650)$ by reproducing the experimental width;
in Ref.~\cite{Guo:2020vmu}, $n = 10$ is adopted for the $Z_{c}(3900)$, $X(4020)$ and $Z_{c}(3985)$ particles.
However, in the Ref.~\cite{Meissner:2015mza}, dedicated to the study of the compositeness of the $\chi_{c1}p$ for the $P_{c}(4450)$,  the upper limit of integration used was  $m_{R}+2\Gamma_{R}$.
The region for $n=2$ is usually thought to be a reasonable cut in the resonance region \cite{Meissner:2015mza,Dias:2021upl}.
For our consideration, we restrict the upper integration limit in the resonance region to $m_{R}+2\Gamma_{R}$ (which comprises the resonance signal as it can  be seen in Fig.~\ref{picflatt}, introduced in Sec.~\ref{sec.201009.3} within the context of a Flatt\'e parametrization).\footnote{We 
  want to stress here that the use of a Flatt\'e parametrization, cf. Sec.~\ref{sec.201009.3}, does not make use of any specific expression for $\Gamma_2$ as a function of $\gamma_2$, like Eq.~\eqref{equ:square}. However, this equation has such a clear physical insight that the results are compatible with those from a Flatt\'e parametrization, as it will be shown in Secs.~\ref{sec.201008.1} and \ref{sec.211126.1} when discussing results.
}

Another aspect to take into account is the RS in which the pole lies because the sign of the momentum of the kaons in the center of mass reference frame has opposite signs in the RSs \Rmnum{2} and \Rmnum{3} at the pole position.
Notice that for the latter RS the kaon momentum has the standard sign in the lower half of the complex energy-plane, corresponding to a $m_R-i\Gamma_1/2-i\Gamma_{2}/2$, while for the former one has instead $m_R-i\Gamma_1/2+i\Gamma_{2}/2$.
 Because this change of sign in the kaon momentum the saturation of the resonance width obtained from the pole position varies, such that $\Gamma_2$ standardly adds to $\Gamma_1$ when the pole lies in the RS \Rmnum{3}, but  $\Gamma_2$ {\it subtracts from} $\Gamma_1$  when the pole lies in the RS \Rmnum{2}. As a result, the total decay width of the $f_{0}(980)$ is then
\begin{equation}
  \label{equ:square1}
\Gamma_{R}=|\gamma_{1}|^{2}\frac{p_{1}(m_{R}^{2})}{8\pi m^{2}_{R}}\pm \frac{|\gamma_{2}|^{2}}{16\pi^{2}} \int_{m_{1}+m_{2}}^{m_{R}+2\Gamma_{R} }dW\frac{p(W^{2})}{W^{2}}\frac{\Gamma_{R}}{(m_{R}-W)^{2}+\Gamma_{R}^2/4}\, ,
\end{equation}
for the pole in the RS \Rmnum{3} or \Rmnum{2}, respectively.

Another interesting consequence of this discussion on the RS in which the pole lies is the simple observation that for a pole in the RS \Rmnum{2} the combination $m_R-i\Gamma_1/2+i\Gamma_2/2$ can be rewritten as $m_R-i(\Gamma_1-2\Gamma_2)/2-i\Gamma_2/2$, so that now the decay width to $K\bar{K}$ appears with the right sign in the resonance propagator for an interpretation as a decay width, while the decay width to the lighter channel is $\Gamma_1-2\Gamma_2$. This result can also be obtained in a more straightforward mathematical way in terms of  the branching ratio to the first channel, $\rx$, by noticing that
\begin{align}
\label{211123.2}
\rx&=1-\frac{\Gamma_2}{\Gamma_R}=\frac{\Gamma_R-\Gamma_2}{\Gamma_R}=\frac{\Gamma_1-2\Gamma_2}{\Gamma_R}\,,~\text{RS \Rmnum{2}}~,
\end{align}
where we have used again that $\Gamma_R=\Gamma_1-\Gamma_2$. However, for a pole in the RS \Rmnum{3} one has the standard result
\begin{align}
\label{211123.1}
\rx&=1-\frac{\Gamma_2}{\Gamma_R}=\frac{\Gamma_1}{\Gamma_R}\,,~\text{RS \Rmnum{3}}~.
\end{align}

Nonetheless, in what follows, we keep the usual notation of directly calling $\Gamma_i$ as decay widths, though for $i=1$ and the pole in the RS \Rmnum{2} the actual decay width to the lighter channel does not coincide with $\Gamma_1$,  as just discussed. Because of this reason we denote by $\Gamma_{\pi\pi}$ or $\Gamma(f_0(980)\to\pi\pi)$ the physical decay width of the $f_0(980)$ to $\pi\pi$, and similarly we use $\Gamma_{\pi\eta}$ or $\Gamma(a_0(980)\to\pi\eta)$ for the  physical partial-decay width of the $a_0(980)$ to the lighter channel. For $K\bar{K}$ we can use indistinctly $\Gamma_2$ or $\Gamma_{K\bar{K}}$ since they coincide. They have the same meaning as $\Gamma(f_0(980)\to K\bar{K})$ or $\Gamma(a_0(980)\to K\bar{K})$ in a clear notation. These points are further elaborated when considering a  Flatt\'e parametrization for the $f_0(980)$ and $a_0(980)$ resonances in Sec.~\ref{sec.201009.3}.

Combining Eq.~(\ref{equ:square2}) and Eq.~(\ref{equ:square1}) allows us to solve $|\gamma_{1}|$ and $|\gamma_{2}|$ in terms of the total compositeness and width. Then we can obtain the partial-decay width $\Gamma_i$ and individual compositeness coefficient $X_{i}$ for each channel.\footnote{The fact that the width of the $f_0(980)$ is substantially larger than the difference between its mass and the $K\bar{K}$ threshold allows to apply in a reasonable way the standard formula Eq.~\eqref{equ1} for the partial compositeness $X_2$, even if the $f_0(980)$ pole lies in the 2nd (3rd) RS above (below) the {\it nearby} $K\bar{K}$ threshold. The mass distribution of the resonance smooths the sharp condition, alluded after Eq.~\eqref{eq:momentum}, on the relative position between the resonance mass and the threshold of $K\bar{K}$ when their distance is much smaller than the $f_0(980)$ width.\label{foot.211008.1}}
We consider several choices for $X$ in Eq.~\eqref{equ:square2}, typically from 0.2 up to 0.8 in steps of 0.2.

The information for the branching ratio  $r_{\rm exp}=\Gamma(f_{0}(980)\rightarrow\pi\pi)/[\Gamma(f_{0}(980)\rightarrow \pi\pi)+\Gamma(f_{0}(980)\rightarrow K\bar{K})]$  can also be used together with the total width $\Gamma_R$ to fix $|\gamma_1|$ and $|\gamma_2|$.
The results following one way or the other are organized in Secs.~\ref{sec.201008.1} and \ref{sec.211126.1}, respectively.
They are also applied in analogous way to the isovector scalar $a_0(980)$ involving the scattering channels $\pi\eta$(1) and $K\bar{K}$(2), with $\rx=\Gamma(a_{0}(980)\rightarrow\pi\eta)/[\Gamma(a_{0}(980)\rightarrow \pi\eta)+\Gamma(a_{0}(980)\rightarrow K\bar{K})]$ then.

\section{Flatt\'e parametrization and the spectral density of a bare state}
\label{sec.201009.3}

One disadvantage of the approach followed in Sec.~\ref{sec.211010.1} is the necessity to assume a value of $n$ for the upper limit of integration in Eq.~\eqref{equ:square} for evaluating the partial-decay width of the resonance into $K\bar{K}$, that is, $\Gamma_2$.
This can be overcome by using a Flatt\'e parametrization \cite{Flatte:1976xu}, without increasing the number of the input parameters needed to calculate the couplings $|\ga_i|$, partial-decay widths $\Gamma_i$ and partial compositeness coefficients $X_i$.
Since the $f_0(980)$ and $a_0(980)$ lie very close to the $K\bar{K}$ threshold  a Flatt\'e parametrization is then especially suitable \cite{Baru:2003qq}.\footnote{Let us notice that the method of Sec.~\ref{sec.211010.1}, based on the saturation of the total width and compositeness of the resonance, can also be applied  to resonances not necessarily lying near a main threshold, like wider or heavier ones.}

 As discussed in Refs.~\cite{Baru:2010ww,Kang:2016jxw}, there is another limitation in the use of a Flatt\'e parametrization as it assumes that the corresponding $K\bar{K}$ partial-wave amplitude has no zero in the near-threshold region.
We assume that this is the case and proceed with the rather intuitive picture offered by a Flatt\'e parametrization of dressing  a bare resonance propagator, $1/D(E)$,  by the self energy due to the intermediate channels 1 and 2,
\begin{align}
\label{210703.1}
D(E)&=E-E_f+i\frac{\widetilde{\Gamma}_1}{2}+\frac{i}{2}g_2\sqrt{m_K E}~.
\end{align}
Here $E$ is the total center of mass energy measured with respect to the two-kaon threshold, $E\equiv \sqrt{s}-2m_K$,
$E_f$ is the bare mass of the resonance plus the contributions at around the $K\bar{K}$ threshold from the real parts  (which are taken as constants) of the meson-meson loops   contributing to the resonance self energy. In addition, $g_i$ is the {\it bare} coupling squared of the resonance to the $i_{\rm th}$ channel,
such that the {\it bare} width $\widetilde\Gamma_1$ to channel 1 is written in terms of $g_1$ as
\begin{align}
\label{211009.2}
\widetilde{\Gamma}_1&=\frac{p_1(m_R)g_1}{8\pi m_R^2}~.
\end{align}
The pole position in the variable $E$ is called $E_R=M_R-i\Gamma_R/2$, {\it with $M_R$ the mass of the resonance with respect to $2m_K$}, $M_R=m_R-2m_K$.

The Flatt\'e parametrization contains as free parameters $E_f$, $\widetilde{\Gamma}_1$ and $g_2$ that can be fixed in terms of the mass
and width of the resonance, that is, by knowing its pole position, and from the knowledge either of the branching ratio $r_{\rm exp}$ to the lighter channel  or the total compositeness $X$.

 To calculate the resonance pole position we must look for the zeroes of
Eq.~\eqref{210703.1}, $D(E_R)=0$,
\begin{align}
\label{210703.2}
E_R-E_f+{\frac{i}{2}}\,\widetilde{\Gamma}_1=-\frac{i}{2}g_2\sqrt{m_K E_R}~.
\end{align}
Taking the square in both sides of the previous expression and solving the resulting quadratic algebraic equation,
we then have the following solutions for the roots
\begin{align}
\label{210703.4}
E_R&=E_f-\frac{1}{8}m_Kg_2^2-\frac{i}{2}\widetilde{\Gamma}_1  +\sigma\sqrt{\frac{m_Kg_2^2}{4}}
\sqrt{\frac{m_Kg_2^2}{16}-E_f+\frac{i}{2}\widetilde{\Gamma}_1}~,
\end{align}
with $\sigma=\pm 1$ in order to keep track of the two different solutions.
Later on we  show that $\sigma=+1(-1)$ corresponds to the pole
lying in the RS \Rmnum{2} (\Rmnum{3}). For the calculation of the square root in the previous equation (taken such that $\Im\sqrt{z}\geq 0$, $z\in \mathbb{C}$) one needs to distinguish two cases according to the sign of
$m_K g_2^2/16-E_f$:
\begin{align}
  \label{210703.5}
 & (\text{i})~\frac{m_Kg_2^2}{16}-E_f > 0~,\\
E_R&=E_f-\frac{i}{2}\widetilde\Gamma_1-\frac{m_K g_2^2}{8}+\frac{\sigma}{2}\sqrt{m_K g_2^2}
\left(\left(\frac{m_Kg_2^2}{16}-E_f\right)^2+\frac{\widetilde\Gamma_1^2}{4}\right)^\frac{1}{4}
\exp{\left(-\frac{i}{2}\arctan\frac{\widetilde\Gamma_1/2}{E_f-m_Kg_2^2/16}\right)}~.\nn
\end{align}
\begin{align}
  \label{210703.6}
&  \text{(ii)}~\frac{m_Kg_2^2}{16}-E_f<0~,\\
  E_R&=E_f-\frac{i}{2}\widetilde\Gamma_1-\frac{m_K g_2^2}{8}+\frac{\sigma}{2}\sqrt{m_Kg_2^2}
  \left(\left(\frac{m_Kg_2^2}{16}-E_f\right)^2+\frac{\widetilde\Gamma_1^2}{4}\right)^\frac{1}{4}
\exp{\frac{i}{2}\left(\pi-\arctan\frac{\widetilde\Gamma_1/2}{E_f-m_Kg_2^2/16}\right)}~.\nn
\end{align}
In what follows we only consider the case (i), because for both (i) and (ii) one obtains the same equations relating $E_f$, $g_2$, and $\widetilde\Gamma_1$ with the inputs $M_R$, $\Gamma_R$, and $r_{\rm exp}$ or $X$.
We introduce the auxiliary angle $\phi$ defined by
\begin{align}
\label{210704.1}
\phi=\arctan\frac{\widetilde\Gamma_1/2}{E_f-m_K g_2^2/16}~.
\end{align}
Therefore, $E_R$ can be written as
\begin{align}
\label{210704.2}
E_R&=E_f-\frac{m_K g_2^2}{8}-\frac{i}{2}\widetilde\Gamma_1+\frac{\sigma}{2}\sqrt{m_Kg_2^2}\left(
\left(E_f-\frac{m_Kg_2^2}{16}\right)^2+\frac{\widetilde\Gamma_1^2}{4}\right)^\frac{1}{4}
(\cos\frac{\phi}{2}-i\sin\frac{\phi}{2})~.
\end{align}
Attending to the real and imaginary parts in this equation we have that
\begin{align}
  \label{210704.3}
\mr&=E_f-\frac{\mk g_2^2}{8}+\sigma \frac{\sqrt{\mk g_2^2}}{2}  \left(\left(E_f-\frac{m_Kg_2^2}{16}\right)^2+\frac{\widetilde\Gamma_1^2}{4}\right)^\frac{1}{4}\cos\frac{\phi}{2}~,\\
\gr&=\widetilde\Gamma_1+\sigma\sqrt{\mk g_2^2}\left(\left(E_f-\frac{\mk g_2^2}{16}\right)^2+\frac{\widetilde\Gamma_1^2}{4}\right)^\frac{1}{4} \sin\frac{\phi}{2}~.\nn
\end{align}
Taking into account the definition of $\phi$, one also has that
\begin{align}
\label{210704.4}
\left(E_f-\frac{\mk g_2^2}{16}\right)^2+\frac{\widetilde\Gamma_1^2}{4}=
\left(E_f-\frac{\mk g_2^2}{16}\right)^2\left(1+\tan^2\phi\right)
=\left(E_f-\frac{\mk g_2^2}{16}\right)^2\frac{1}{\cos^2\phi}~,
\end{align}
and
\begin{align}
\label{210704.6}
E_f-\frac{\mk g_2^2}{16}&=\frac{\widetilde\Gamma_1}{2}\cot\phi~.
\end{align}
Substituting these two equalities into Eq.~\eqref{210704.3}, with $\phi<0$ for case (i), the latter equation becomes
\begin{align}
\label{210704.7}
\mr&=
-\frac{\mk g_2^2}{16}+\frac{\widetilde\Gamma_1}{2}\cot\phi+\frac{\sigma}{4}\sqrt{\mk g_2^2\widetilde\Gamma_1|\cot\frac{\phi}{2}|}~,\\
\gr&=
\widetilde\Gamma_1- \frac{\sigma}{2}\sqrt{\mk g_2^2\widetilde\Gamma_1 |\tan\frac{\phi}{2}|}~.\nn
\end{align}
From this last equation it follows that
\begin{align}
\label{210704.9}
\frac{\sigma}{2}\sqrt{\mk g_2^2\widetilde{\Gamma}_1}&= (\widetilde{\Gamma}_1-\gr)\sqrt{|\cot\frac{\phi}{2}|}~.
\end{align}
When this is taken into Eq.~\eqref{210704.7}  we can write $\mr$ as
\begin{align}
\label{210705.1b}
M_R&=\frac{\Gamma_R^2}{4\widetilde{\Gamma}_1}\cot\frac{\phi}{2}\left[1-\left(\frac{\widetilde{\Gamma}_1}{\Gamma_R}\tan\frac{\phi}{2}\right)^2\right]~,\\
\gr&=\widetilde{\Gamma}_1- \frac{\sigma}{2}\sqrt{\mk g_2^2\widetilde{\Gamma}_1 |\tan\frac{\phi}{2}|}~.\nn
\end{align}
The equation for $\mr$ is of the form,
\begin{align}
\label{210801.6}
&x-\frac{1}{x}=\frac{4 M_R}{\Gamma_R}~,\\
&x\equiv \frac{\Gamma_R}{\widetilde{\Gamma}_1}\cot\frac{\phi}{2}<0~,\nn
\end{align}
and its solution for $x<0$ is
\begin{align}
\label{210801.7}
x&=\frac{2 M_R}{\Gamma_R}-\sqrt{1+\left(\frac{2M_R}{\Gamma_R}\right)^2}~.
\end{align}
After this is substituted  in the expression for  $\Gamma_R$ in Eq.~\eqref{210705.1b}, we can isolate the bare partial-decay width to the first channel,
and then the bare branching ratio $r\equiv \widetilde{\Gamma}_1/\Gamma_R$ is given by
\begin{align}
\label{210801.9}
r&=1+\frac{\sigma g_2\sqrt{m_K}/2}{u-2M_R}~,\\
u&\equiv (4M_R^2+\Gamma_R^2)^{1/2}~.\nn
\end{align}
Once we know $r$ we can also determine $\cot\phi/2$ by using the definition of $x$ and its  solution  in Eq.~\eqref{210801.7},
\begin{align}
\label{210801.10}
\cot\frac{\phi}{2}&=r x=
\frac{1}{2\Gamma_R}
\left(4M_R-2u
-\sigma g_2 \sqrt{m_K (u-2M_R)}\right)~.
\end{align}

Let us denote by $\beta$ the residue of $1/D(E)$ at the resonance pole,
\begin{align}
\label{210913.2}
\beta&= \left|\lim_{E\to E_R}\frac{E-E_R}{D(E)}\right|=\left|\frac{1}{1+\frac{i g_2}{4}\sqrt{\frac{\mk}{E_R}} }\right|=\frac{\sqrt{8u}}{\left(g_2^2 m_K+8 u+  4 \sigma g_2 \sqrt{m_K(u-2M_R)}\right)^{1/2}}~.
\end{align}
This expression can be obtained by substituting $\sqrt{E_R}$ from Eq.~\eqref{210703.2},
the relation between $E_f$ and $\tan\phi$, cf. Eq.~\eqref{210704.6}, and finally
the expression for $\tan\phi$ from that of $\cot\phi/2$ given in Eq.~\eqref{210801.10}.

The {\it renormalized or  dressed} coupling squared $|\gamma_i^2|$ is related to the bare one $g_i$ by evaluating the residue of the elastic
scattering amplitudes for channel $i$, $g_i/D(E)$. In terms of $\beta$, we have the result
\begin{align}
\label{210801.12}
|\gamma_1|^2=g_1\beta~,\\
|\gamma_2|^2=32\pi m_K^2g_2\beta~,\nn
\end{align}
with the numerical factor in front of $g_2\beta$
needed for having the same normalization as in Sec.~\ref{sec.211010.1}.

Up to our knowledge the difference between the bare and dressed couplings in a Flatt\'e parametrization has not been clearly discussed before in the literature, and it has important implications. E.g.
this is  one of the reasons why  the values for $\widetilde{\Gamma}_{\pi\pi}$ collected in the Table~2 of Ref.~\cite{Baru:2003qq} for the $f_0(980)$ are
typically much bigger than 100~MeV.
A similar comment can also be made for most of the entries of $\widetilde{\Gamma}_{\pi\eta}$ in Table~1 of the same reference regarding the $a_0(980)$.\footnote{The other reason applies to those poles in the RS II because then the physical partial-decay width to the lighter channel is $\Gamma_1-2\Gamma_2$, which is smaller than $\Gamma_1$, cf. Eq.~\eqref{211123.2} and discussions below in this section.}   Indeed, this can  be a source for confusion in the literature. In this respect, we notice that Ref.~\cite{GarciaMartin:2011jx} compares its $\pi\pi$ $S$-wave residue with bare couplings used in energy-dependent-width Breit-Wigner or Flatt\'e parametrizations without  considering the actual residue at the resonance pole position of the parametrization.

In order to obtain $g_2$ we need another input, for which we take either the physical branching ratio $r_{\rm exp}$ or the total compositeness $X$.
For the latter case we need then the expression for calculating $X_1$ and $X_2$, with $X=X_1+X_2$. Recalling Eq.~\eqref{equ1} we have for $X_1$,
\begin{align}
  \label{211012.3}
  X_1&=\gamma_1^2 \left|\frac{\partial G_1}{\partial s}\right|_{s=s_R}
=\frac{8\pi m_R^2\Gamma_R}{p_1(m_R)} r \beta \left|\frac{\partial G_1}{\partial s}\right|_{s=s_R}~,
\end{align}
with $r$ and $\beta$ given in terms of $g_2$ and the pole parameters in Eqs.~\eqref{210801.9} and \eqref{210913.2}, respectively.
For the calculation of $X_2$ a simpler algebraic formula can be obtained if we use nonrelativistic kinematics for the calculation of the derivative of $\partial G_2(s)/\partial s$ at $s=s_R$, taking advantage of the fact that pole lies in the vicinity of the $K\bar{K}$ threshold. Then, $G_2(s)$ is a constant plus $-i \sqrt{m_K E}/(16\pi m_K)+{\cal O}(E/m_K)$, and its derivative with respect to $s$ is
\begin{align}
\label{211106.1}
  \left.\frac{\partial G_2(s)}{\partial s}\right|_{s_R}=
  \frac{-i}{128\pi m_K^{3/2}\sqrt{E_R}}+{\cal O}(1)~.
\end{align}
We then multiply this derivative by $32\pi m_K^2 g_2 \beta$ and the final expression that results is
\begin{align}
\label{210801.18}
X_2&=\frac{\sqrt{m_K}g_2}{\left(g_2^2m_K+8u+4\sigma g_2 \sqrt{m_K(u-2M_R)}\right)^{\frac{1}{2}}}~.
\end{align}
The equation that is needed to be solved to obtain $g_2$ given $X$ is $X_1+X_2=X$.
Nonetheless, in the numerical results shown below we calculate $X_2$ making use of relativistic kinematics, with differences of around a $10\% - 15$\% compared with the values obtained when using the nonrelativistic Eq.~\eqref{210801.18}.

When $\rx$ is the input taken one has to distinguish between whether the pole lies in the RS \Rmnum{2} or RS \Rmnum{3}, due to the change of sign in the analytical extrapolation of $\sqrt{E}$ in $D(E)$, Eq.~\eqref{210703.1}, needed to reach the pole position.
For the pole in the RS \Rmnum{3}, we have the following straightforward relation between the physical $r_{\rm exp}$ and the bare $r$,
\begin{align}
\label{211012.1}
r_{\rm exp}&=r \beta=\frac{\sqrt{2u}\left(2\sqrt{u-2M_R}+g_2\sigma\sqrt{m_K}\right)}{\sqrt{u-2M_R}\left(g_2^2 m_K+8u+4\sigma g_2 \sqrt{m_K(u-2M_R)}\right)^{1/2}}~.
\end{align}
This is a quadratic equation for $g_2$ that can be easily solved. By requiring that $g_2=0$ for $\rx=1$ (which implies that there is no resonance decay at all to the $K\bar{K}$ channel) then there is only one acceptable solution given by
\begin{align}
\label{211012.2}
g_2&=\frac{2\left(-\Gamma_R r_{\exp}+\sqrt{u-2M_R}\sqrt{2M_R r_{\rm exp}^2+(2-r_{\rm exp}^2)u}\right)}{\sqrt{m_K\left(2M_R r_{\rm exp}^2+(2-r_{\rm exp}^2)u\right)}}~.
\end{align}

For a pole in the RS \Rmnum{2}, when moving to the complex $E$-plane with negative imaginary part  the square root $\sqrt[II]{E}$ has a positive imaginary part and negative real one, changing sign with respect to that when calculated in the RS \Rmnum{3}, because there $\sqrt[III]{E}=-\sqrt[II]{E}$ (the momentum $p_2$ changes sign between the two sheets).
As a result, it is indeed the case that the last term in Eq.~\eqref{210703.1}, responsible for the width to $K\bar{K}$, does not add to but subtract from $\widetilde{\Gamma}_1$.
Then, for a physical interpretation of the different terms in this equation it is convenient to rewrite it as
\begin{align}
\label{211109.1}
D(E)&=E-E_f+\frac{i}{2}\left[\widetilde{\Gamma}_1+2g_2\sqrt[II]{m_K E}\right]-\frac{i}{2}g_2\sqrt[II]{m_K E}\\
&=E-E_f+\frac{i}{2}\left[\widetilde{\Gamma}_1-2g_2\sqrt[III]{m_K E}\right]+\frac{i}{2}g_2\sqrt[III]{m_K E}
~.\nn
\end{align}
 Let us notice that the real part of $\sqrt[III]{M_R-i\,\Gamma_R/2}$ is positive and then
the imaginary part of the last term in Eq.~\eqref{211109.1} appears with the same sign as $\widetilde{\Gamma}_1$.
From this equation we then deduce that, once the bare couplings are dressed, cf. Eq.~\eqref{210801.12}, the partial-decay width to the lighter channel really observed in an experiment is not $\Gamma_1$ but $\Gamma_1-2\Gamma_2$, while the total decay width $\Gamma_R$ is  $\Gamma_1-\Gamma_2$ for the RS \Rmnum{2} case. Therefore, when
taking $r_{\rm exp}$ as input for a pole in the RS \Rmnum{2},
\begin{align}
\label{211110.1}
r_{\rm exp}&=\frac{\Gamma_1}{\Gamma_R}-2\frac{\Gamma_2}{\Gamma_R}=r\beta-2(1-\rx)
\end{align}
from where  the extra equation to be taken into account is:
\begin{align}
\label{211110.2}
2-r_{\rm exp}&=r\beta=\frac{\sqrt{2u}\left(2\sqrt{u-2M_R}+g_2\sigma\sqrt{m_K}\right)}{\sqrt{u-2M_R}\left(g_2^2 m_K+8u+4\sigma g_2 \sqrt{m_K(u-2M_R)}\right)^{1/2}}~,
\end{align}
instead of the straight Eq.~\eqref{211012.1} for a pole in the RS \Rmnum{3}.
The valid solution in this case, the one that gives $g_2=0$ for $\rx=1$, is obtained from Eq.~\eqref{211012.2} by simultaneously multiplying its right-hand side by a minus sign and replacing $\rx$ for $2-\rx$.

As a result of this analysis  we then expect that $\Gamma_1>\Gamma_R$ with values in the interval $[\Gamma_R,2\Gamma_R]$ for the resonance when its pole lies in the RS \Rmnum{2}.
To illustrate this point, let us take the residues given in the Refs.~\cite{GarciaMartin:2011jx,Guo:2012yt} for the pole positions of the $f_0(980)$ in Eqs.~\eqref{equ22} and \eqref{211107.1}, respectively, and evaluate $\Gamma_1/\Gamma_R$ making use of Eq.~\eqref{equ33}.
The values of the residues are $|\gamma_1|=2.3\pm 0.2$~GeV \cite{GarciaMartin:2011jx} and $|\gamma_1|=1.80\pm 0.25$~GeV \cite{Guo:2012yt}. Propagating the errors in $m_R$, $\Gamma_R$ and $|\gamma_1|$ from these references, the values that we obtain are
\begin{align}
\label{211111.1}
  \frac{\Gamma_1}{\Gamma_R}&=2.0\pm 0.9\,,~\text{Eq.~\eqref{equ22}--Ref.~\cite{GarciaMartin:2011jx}}~,\\
  \frac{\Gamma_1}{\Gamma_R}&=1.1\pm 0.6\,,~\text{Eq.~\eqref{211107.1}--Ref.~\cite{Guo:2012yt}}~.\nn
\end{align}
This more detailed discussion based on the use of the Flatt\'e parametrization extends the same topic already discussed in Sec.~\ref{sec.211010.1} for the interpretation of the $\Gamma_i$'s and their connection with the experimental decay widths for the case of a pole lying in the RS \Rmnum{2}.

It is worth stressing  that when $X$ is taken as input the procedure explained above, Eqs.~\eqref{211012.3}-\eqref{210801.18}, allows to calculate $\Gamma_1/\Gamma_R$ without relying on its connection to $\rx$, and the resulting values, which we discuss below, are perfectly compatible with the picture just explained.

An interesting outcome of our study is that by combining outputs from different sources we are able to constrain further the knowledge on the nature of the resonances $f_0(980)$ and $a_0(980)$ than by just relying on one reference. This is clear by considering the huge uncertainty in the decay branching ratios $\Gamma_1/\Gamma_R$ deduced in Eq.~\eqref{211111.1} from Refs.~\cite{GarciaMartin:2011jx} and \cite{Guo:2012yt}. However, later on we will use these same references for the mass and width and then take the branching ratio from other studies so that one can then conclude a tighter information on the nature for the $f_0(980)$. Related to this, we will be able to deduce values for the dressed coupling squared $|\gamma_2|^2$ which cannot be provided by the formalism of Refs.~\cite{GarciaMartin:2011jx,Guo:2012yt}.  Similar remarks also hold for the $a_0(980)$.

Finally, let us now show the relationship between the sign $\sigma$ and the RS in which the pole lies.
For that, we isolate $\sqrt{E_R}$ from Eq.~\eqref{210703.2}, which can then be written as
\begin{align}
\label{210801.14}
\sqrt{E_R}&=\frac{2i}{g_2 \sqrt{m_K}}(M_R-E_f)+\frac{\Gamma_R-\widetilde{\Gamma}_1}{g_2 \sqrt{m_K}}~.
\end{align}
This equation tells us that if $M_R-E_f>0$ the pole lies in the RS \Rmnum{2}, since then $\Im \sqrt{E_R}>0$ and, conversely, if $M_R-E_f<0$ the pole is located in the RS \Rmnum{3}.
Now, we consider Eq.~\eqref{210704.3} which clearly implies that if $\sigma=-1$ then $M_R-E_f<0$, corresponding to the RS \Rmnum 3.
For $\sigma=+1$ a more careful treatment is needed because the sign of $M_R-E_f$ depends on the relative sign between the last two terms in the right-hand side of Eq.~\eqref{210704.3}.
One can straightforwardly show that the absolute value of the last term is bigger than $m_Kg_2^2/8$ by squaring and subtracting them.
In the process one has to relate $E_f$ with $\tan\phi$, Eq.~\eqref{210704.6}, and use the expression for $\cot\phi/2$ given in Eq.~\eqref{210801.10}.
As a result $M_R-E_f>0$ and the pole lies in the RS \Rmnum 2 for $\sigma=+1$.

We give  the results obtained with the present formalism based on the use of the Flatt\'e parametrization, distinguishing between when $X$ or $r_{\rm exp}$ are taken as inputs in Secs.~\ref{sec.201008.1} and \ref{sec.211126.1}, respectively.

\subsection{Spectral density and its integration}
\label{sec.211106.2}

Here we use the spectral density function $\omega(E)$ of a near-threshold resonance, in our case either $f_0(980)$ or $a_0(980)$, as a way to calculate the compositeness of the  meson-meson states in these resonances.
We follow the formalism of Ref.~\cite{Baru:2003qq} to which we refer for further details.
There the spectral density function $\omega(E)$ is introduced, and it provides the probability distribution function in energy for finding a bare elementary state in the continuum \cite{Bogdanova:1991zz}.
As a result, its integration  around the $K\bar{K}$ threshold comprising the resonance signal, which we call $W_R$,  is the probability for finding the bare state. Namely, $W_R$ is calculated as \cite{Baru:2003qq}
\begin{align}
\label{211013.1}
  W_R&=\int_{-\Delta}^{+\Delta} dE\, \omega(E)~,\\
  \omega(E)&=\frac{1}{2\pi}
\frac{\widetilde{\Gamma}_1+g_2\sqrt{m_KE}\theta(E)}{\left(E-E_f-\frac{1}{2}g_2\sqrt{-m_KE}\theta(-E)\right)^2+\frac{1}{4}\left(\widetilde{\Gamma}_1+g_2\sqrt{m_KE}\theta(E)\right)^2}~,\nn
\end{align}
with $\theta(E)$ the Heaviside step function.
In Ref.~\cite{Baru:2003qq} the parameter $\Delta$ was chosen to be $50$~MeV but, since an important dependence on $\Delta$ is observed on the final value of $W_R$, we prefer to present the results for $W_R$ as a function of $\Delta$.

\subsection{Reinterpretation of the method of Sec.~\ref{sec.211010.1} for poles in the Riemann sheet \Rmnum{2}}
\label{sec.211110.1}

By the application of the Flatt\'e parametrization it has been clear that for a near-threshold pole in coupled channels lying in the RS \Rmnum{2} it is necessary to change the interpretation of $\Gamma_1$.
We have seen from the last line in Eq.~\eqref{211109.1} that the partial-decay width into the lighter channel is not directly $\Gamma_1$ but $\Gamma_1-2\Gamma_2$, and that the total width from the pole position should be compared with $\Gamma_1-\Gamma_2$.

Therefore, for a pole in the RS \Rmnum{2} near the heavier threshold the equation for the saturation of the total width, cf. Eq.~\eqref{equ:square1},  reads
\begin{align}
  \label{211111.3}
  \Gamma_{R}=|\gamma_{1}|^{2}\frac{p_{1}(m_{R}^{2})}{8\pi m^{2}_{R}}
  -
\frac{|\gamma_{2}|^{2}}{16\pi^{2}} \int_{m_{1}+m_{2}}^{m_{R}+2\Gamma_{R} }dW\frac{p(W^{2})}{W^{2}}\frac{\Gamma_{R}}{(m_{R}-W)^{2}+\Gamma_{R}^2/4}~,
\end{align}
with no change in Eq.~\eqref{equ:square2} for saturating the total compositeness $X$. The latter equation is also reproduced here
\begin{equation}
  \label{211111.4}
X=|\gamma_{1}|^{2}\Big{|} \frac{\partial G_{1}(s)}{\partial s}\Big{|}_{s=s_{R}}+|\gamma_{2}|^{2}\Big{|} \frac{\partial G_{2}(s)}{\partial s}\Big{|}_{s=s_{R}}\,.
\end{equation}
However, in applications up to now of these ideas instead of the minus in Eq.~\eqref{211111.3} a ``standard'' plus sign is placed in front of $|\gamma_2|^2$. Namely,
\begin{align}
  \label{211128.1}
  \Gamma_{R}=|\gamma_{1}|^{2}\frac{p_{1}(m_{R}^{2})}{8\pi m^{2}_{R}}
  +\frac{|\gamma_{2}|^{2}}{16\pi^{2}} \int_{m_{1}+m_{2}}^{m_{R}+2\Gamma_{R} }dW\frac{p(W^{2})}{W^{2}}\frac{\Gamma_{R}}{(m_{R}-W)^{2}+\Gamma_{R}^2/4}~,
\end{align}

At the practical level as long as $X_1\ll X_2$ the only change is in the value of $|\gamma_1|$, because the use of Eq.~\eqref{211128.1} would provide an ``effective'' value of this coupling by reabsorbing the effect of subtracting $-2\Gamma_2$ in Eq.~\eqref{211111.3}.
This is clear because then $X$ is saturated almost completely by $X_2$ and this quantity alone fixes $|\gamma_2|$.
The use of Eq.~\eqref{211111.3} allows to determine $|\gamma_1|$, which is certainly larger than the ``effective'' one deduced  by fulfilling Eq.~\eqref{211128.1} with an intermediate plus sign.
However, the values for $X_2$, $\Gamma_2$ and the physical partial-decay width to channel 1 almost do not change, which are typically the most important pieces of information to infer about the nature of the resonances.

Now, if $\rx$ is taken as input the equations that one has to fulfill are Eq.~\eqref{211111.3} and
\begin{align}
\label{211111.5}
2-\rx=|\gamma_{1}|^{2}\frac{p_{1}(m_{R}^{2})}{8\pi m^{2}_{R}\Gamma_R}~,
\end{align}
which are apparently very  different to the ``standard'' Eq.~\eqref{211128.1} and
\begin{align}
  \label{211128.2}
  \rx=\frac{\Gamma_1}{\Gamma_R}~,
\end{align}
already  written for a pole in the RS \Rmnum{3} in Sec.~\ref{sec.211010.1}.
However, they basically provide again  the same results for $\Gamma_2$, $X_2$, and the physical partial-decay width to channel 1, while the use of Eqs.~\eqref{211128.1} and \eqref{211128.2} provides the already mentioned ``effective'' value for $|\gamma_1|$.

In order to see it, let us divide  Eq.~\eqref{211111.3} by $\Gamma_R$ and    write the following equations equivalent to Eqs.~\eqref{211111.3} and \eqref{211111.5},
\begin{align}
  \label{211111.6}
1&=\frac{\Gamma_1}{\Gamma_R}-\frac{\Gamma_2}{\Gamma_R}~,\\
2-\rx&=\frac{\Gamma_1}{\Gamma_R}~.\nn
\end{align}
Subtracting the first  line from the second one we then have
\begin{align}
  \label{211111.7}
  1-\rx&=  \frac{\Gamma_2}{\Gamma_R}~,
\end{align}
which fixes $|\gamma_2|$ as if Eq.~\eqref{211128.1} were used.
Afterwards, Eq.~\eqref{211111.3} is employed to calculate $|\gamma_1|$, while the use of Eq.~\eqref{211128.1} would provide the so-called ``effective'' value for $|\gamma_1|$, which would be smaller than the one obtained from Eq.~\eqref{211111.3}.

\section{Results and discussions using the total compositeness as input}\label{section1}
\label{sec.201008.1}

Here we apply the formalism derived in Secs.~\ref{sec.211010.1} and \ref{sec.201009.3} to study the nature of the resonances $f_0(980)$ and $a_0(980)$. Subsequently, the former method based on the saturation of $\Gamma_R$ and $X$ is denoted by S, while the latter one based on the use of a Flatt\'e parametrization is called $F$. For each resonance pole we first apply the method S and then F.

\subsection{The $f_0(980)$ resonance}
\label{sec.201009.1}
Assuming given values for the total compositeness $X$ of the $f_{0}(980)$, varying it
from $0.2$ to $1.0$ in steps of 0.2, 
we obtain the couplings, partial-decay widths, and compositeness coefficients by solving Eq.~(\ref{equ:square2}) and Eq.~(\ref{equ:square1}). We indicate that it is not possible to find solutions for the parameters to reproduce  $X=0$, because then the couplings would be zero which is in contradiction of having a finite width.
As already mentioned, we  consider the  RS \Rmnum{2} where $s_R$ of Eq.~\eqref{equ22} lies \cite{GarciaMartin:2011jx} and the  results calculated are shown in Table~\ref{tabff1}.
One can observe from this table that the compositeness coefficient $X_{2}$ is always much larger than $X_{1}$,
which means that the $K\bar{K}$ channel plays a much more important role than the $\pi\pi$ one in the structure of  $f_{0}(980)$.
Related to this point, the $f_{0}(980)$ couples much more strongly to $K\bar{K}$ than to $\pi\pi$, in agreement with the fact that $f_{0}(980)$ sits very close to the $K\bar{K}$ threshold.
We also observe from Table~\ref{tabff1} that as the total compositeness $X$ increases
the physical partial-decay width to $\pi\pi$ (here, for a RSII pole, $\Gamma_{\pi\pi}=\Gamma_1-2\Gamma_2$) decreases, while that to $K\bar{K}$ becomes larger.
Similarly, this fact has also been found in other heavy-quark resonances with  open and  near-threshold channels \cite{Du:2021bgb,Guo:2020vmu,Guo:2020pvt,Guo:2019kdc}.
In these regards, it follows from Table~\ref{tabff1} that
 $\Gamma_{K\bar{K}}>\Gamma_{\pi\pi}$ for $X\gtrsim 0.6$,\footnote{This can be easily seen because $\Gamma_{K\bar{K}}>25$~MeV.} but as $X$ decreases $\Gamma_{\pi\pi}$  becomes larger than $\Gamma_{K\bar{K}}$.

For estimating the errors associated to the input values of the resonance pole mass and width we proceed in this work similarly as done in Ref.~\cite{Kang:2016ezb}. Then, we discretize the inputs, $m_R$ and $\Gamma_R$ from the pole position at several points within one standard deviation region from the central values to generate  a data grid.
For each of the points in the grid we proceed to calculate the different outputs so that
their central values correspond to the mean values and
the errors to the square root of the variances. We have also checked that this procedure
is (of course) stable if the number of points in the
grid is increased.  For the other input $X$ the variation in the results calculated for the different values of $X$  provides an estimate of this source of uncertainty.

\begin{table}[!htbp]
\begin{center}
  \caption{{\small Method S applied to the resonance $f_0(980)$ with pole position in the RS \Rmnum{2} (column 2) from Ref.~\cite{GarciaMartin:2011jx}, Eq.~\eqref{equ22}: The couplings $|\ga_i|$ (columns 3, 4), corresponding partial-decay widths $\Gamma_{i}$ (columns 5, 6), and individual compositeness coefficients $X_{i}$ (columns 7, 8) are calculated for $X$ taking values from 0.2 up to 1.0 in steps of 0.2} (column 1).
  }
\begin{tabular}{|cc|cccccc|}
\Xhline{1pt} ~~$X~~~$&~~~$\text{RS}$&~~~$|\gamma_{\pi\pi}|(\text{GeV})$~~~&$|\gamma_{K\bar{K}}|(\text{GeV})$~~~&$\Gamma_{1}(\text{MeV})$&~~~$\Gamma_{2}(\text{MeV})$&~~~$X_{\pi\pi}$&~~~$X_{K\bar{K}}$~~~\\
\Xhline{1pt} \multirow{1}*{1.0}&~~~$\text{\Rmnum{2}}$&~~~$2.37\pm0.21$~~~&$5.21\pm0.26$~~~&$108.3\pm18.9$&~~~$54.3\pm10.6$&~~~$0.042\pm0.007$&~~~$0.958\pm0.007$~~~\\
\Xhline{1pt} \multirow{1}*{0.8}&~~~$\text{\Rmnum{2}}$&~~~$2.24\pm0.20$~~~&$4.65\pm0.23$~~~&$97.2\pm16.9$&~~~$43.2\pm8.4$&~~~$0.038\pm0.007$&~~~$0.762\pm0.007$~~~\\
\Xhline{1pt} 		\multirow{1}*{0.6}&~~~$\text{\Rmnum{2}}$&~~~$2.11\pm0.19$~~~&$4.01\pm0.19$~~~&$86.1\pm14.8$&~~~$32.1\pm6.2$&~~~$0.033\pm0.006$&~~~$0.567\pm0.006$~~~\\
\Xhline{1pt} \multirow{1}*{0.4}&~~~$\text{\Rmnum{2}}$&~~~$1.97\pm0.17$~~~&$3.24\pm0.15$~~~&$75.0\pm12.9$&~~~$21.0\pm4.0$&~~~$0.029\pm0.005$&~~~$0.371\pm0.005$~~~\\
\Xhline{1pt}
\multirow{1}*{0.2}&~~~$\text{\Rmnum{2}}$&~~~$1.82\pm0.16$~~~&$2.23\pm0.09$~~~&$63.9\pm11.0$&~~~~$9.9\pm1.8$&~~~$0.025\pm0.004$&~~~$0.175\pm0.004$~~~\\
\Xhline{1pt}
\end{tabular}
\label{tabff1}
\end{center}
\end{table}

We now consider the application of the method F,  take  $X=0.2$ to 0.8 
in steps of 0.2, and calculate the parameters   characterizing the Flatt\'e formula, Eq.~\eqref{210703.1}, that is, the bare width to the lighter channel $\widetilde{\Gamma}_1$, the bare coupling squared $g_2$ to $K\bar{K}$ and $E_f$.
As typical outputs we provide $\Gamma_1$, $\Gamma_2$, $X_1$ and $X_2$.
For the pole position of the $f_0(980)$ we take  Eq.~\eqref{equ22}, with the results given in Table~\ref{tabflatt1}.
For this method the extreme value $X=1$ cannot be reproduced, while $X=0$ is in contradiction with having a finite width as for the method S.

Comparing the values of $X_2$ for the $f_0(980)$ between Table~\ref{tabflatt1} and the previous Table~\ref{tabff1} we see a good agreement, with the difference in the central values affecting the third decimal figure. This close agreement can be explained, though not completely, by  noticing that $X_2\gg X_1$ so that $X_2\approx X$, which fixes it to lie very close to the total compositeness.  Regarding the $\Gamma_i$ we see that they are very close between both tables for $X=0.2$, 0.4 and 0.6, with larger differences for $X=0.8$, but perfectly compatible within errors for all values of $X$. 
One should stress that the method based on the Flatt\'e parametrization directly provides $\Gamma_i$, since $\Gamma_1=\widetilde{\Gamma}_1\beta$ and $\Gamma_2=\Gamma_1-\Gamma_R$  for a pole in the RS \Rmnum{2}, as it is the case here for the $f_0(980)$ poles in Eqs.~\eqref{equ22} and \eqref{211107.1}.
However, the method S in Sec.~\ref{sec.211010.1} keeps some spurious dependence on the value given to  $n$ for Eq.~\eqref{equ:square}, though a judicious choice gives rise to remarkably compatible results between both methods, as indicated. 

\begin{table}[!htbp]
\begin{center}
\caption{ Method F applied to the resonance $f_0(980)$ with pole position  in the RS $\text{\Rmnum{2}}$ (column 2) from Ref.~\cite{GarciaMartin:2011jx}, Eq.~\eqref{equ22}: The value of $X$ taken as input is given in the first column. We calculate the bare width $\widetilde{\Gamma}_1$ (column 3), the bare coupling $g_2$ (column 4),  $E_{f}$ (column 5), $\Gamma_{1}$ (column 6), $\Gamma_{2}$ (column 7), $X_1$ (column 8),  and $X_{2}$ (column 9). }
\begin{tabular}{|cc|ccccccc|}
\Xhline{1pt}
$X~$&~$\text{RS}$&$\widetilde{\Gamma}_{\pi\pi}$(MeV)&$g_2$&$E_f$(MeV)&$\Gamma_{1}$(MeV)&$\Gamma_{2}$(MeV)&$X_{\pi\pi}$&$X_{K\bar{K}}$\\
\Xhline{1pt}
\multirow{1}*{0.8}&$\text{\Rmnum{2}}$&$~948.3\pm383.1$&\!\!$10.01\pm4.03$&$-389.7\pm210.7$&$~84.4\pm15.3$&$~30.4\pm8.3$&$~0.033\pm0.006$&$~0.767\pm0.006$\\
\Xhline{1pt}
\multirow{1}*{0.6}&$\text{\Rmnum{2}}$&$200.3\pm38.5$&$1.63\pm0.23$&$-57.6\pm19.3$&$~80.4\pm14.4$&$~26.4\pm7.0$&$~0.031\pm0.006$&$~0.569\pm0.006$\\
\Xhline{1pt}
\multirow{1}*{0.4}&$\text{\Rmnum{2}}$&$113.2\pm20.1$&$0.66\pm0.07$&$-20.4\pm8.6$&$~73.2\pm12.9$&$~19.2\pm4.9$&$~0.028\pm0.005$&$~0.372\pm0.005$\\
\Xhline{1pt}
\multirow{1}*{0.2}&~$\text{\Rmnum{2}}$&$75.2\pm13.0$&$0.24\pm0.02$&$-4.2\pm5.3$&$~63.9\pm11.1$&$~9.9\pm2.4$&$~0.025\pm0.004$&$~0.175\pm0.004$\\
\Xhline{1pt}
\end{tabular}
\label{tabflatt1}
\end{center}
\end{table}

 We now move on and consider the $f_0(980)$ pole position from Ref.~\cite{Guo:2012yt} given in Eq.~\eqref{211107.1}, and proceed  similarly as done regarding the pole position in Eq.~\eqref{equ22}. Then,  we reproduce the values of the mass and width of the resonance in Eq.~\eqref{211107.1} together with a value given for $X$. The results obtained are shown  in Table~\ref{tab.211106.1}. 

Comparing Tables~\ref{tabff1} and \ref{tab.211106.1} we observe that all the outputs are rather similar. In particular the resulting values for $X_2$ are almost coincident in both Tables, with $X_2\gg X_1$. However, we notice that while the central value of the width for the pole in Eq.~\eqref{equ22} is smaller than that in the pole of Eq.~\eqref{211107.1}, the calculated $\Gamma_{1}$ and $\Gamma_{2}$ in Table~\ref{tabff1} are larger than those in Table~\ref{tab.211106.1}. This indicates that the cancellation between widths in the difference $\Gamma_{1}-2\Gamma_{2}$ for calculating the actual width to $\pi\pi$ is more important in the pole of Eq.~\eqref{equ22} than for the one in Eq.~\eqref{211107.1}.

\begin{table}[!htbp]
\begin{center}
	\caption{{\small Method S applied to the resonance $f_0(980)$ with pole position  in the RS \Rmnum{2}  from Ref.~\cite{Guo:2012yt}, Eq.~\eqref{211107.1}. The various entries are the same as in Table~\ref{tabff1}.} \label{tab.211106.1}}
	\begin{tabular}{|cc|cccccc|}
		\Xhline{1pt}
		~~$X~~~$&~~~$\text{RS}$&~~~$|\gamma_{\pi\pi}|(\text{GeV})$~~~&$|\gamma_{K\bar{K}}|(\text{GeV})$~~~&$\Gamma_{1}(\text{MeV})$&~~~$\Gamma_{2}(\text{MeV})$&~~~$X_{\pi\pi}$&~~~$X_{K\bar{K}}$~~~\\
		\Xhline{1pt}
\multirow{1}*{1.00}&~~~$\text{\Rmnum{2}}$&~~~$2.18\pm0.28$~~~&$5.47\pm0.34$~~~&$94.3\pm23.3$&~~~~\,$38.3\pm12.7$&~~~$0.037\pm0.009$&~~~$0.963\pm0.009$~~~\\
\Xhline{1pt} 		\multirow{1}*{0.80}&~~~$\text{\Rmnum{2}}$&~~~$2.09\pm0.26$~~~&$4.88\pm0.30$~~~&$86.5\pm20.8$&~~~~\,$30.5\pm10.0$&~~~$0.034\pm0.008$&~~~$0.766\pm0.008$~~~\\
\Xhline{1pt} 		\multirow{1}*{0.60}&~~~$\text{\Rmnum{2}}$&~~~$2.00\pm0.24$~~~&$4.21\pm0.26$~~~&$78.6\pm18.4$&~~~$22.6\pm7.4$&~~~$0.031\pm0.007$&~~~$0.569\pm0.007$~~~\\
		\Xhline{1pt} \multirow{1}*{0.40}&~~~$\text{\Rmnum{2}}$&~~~$1.89\pm0.22$~~~&$3.40\pm0.20$~~~&$70.8\pm16.0$&~~~$14.1\pm4.8$&~~~$0.028\pm0.006$&~~~$0.372\pm0.006$~~~\\
		\Xhline{1pt}
\multirow{1}*{0.20}&~~~$\text{\Rmnum{2}}$&~~~$1.79\pm0.20$~~~&$2.33\pm0.13$~~~&$62.9\pm13.7$&~~~~$6.9\pm2.2$&~~~$0.025\pm0.005$&~~~$0.175\pm0.005$~~~\\
		\Xhline{1pt}
	\end{tabular}
\end{center}
\end{table}

For the method F the results are given in Table~\ref{tab.211107.1}.  We observe that the  values of $X_2$ obtained are very similar to those given in Table~\ref{tab.211106.1} for the same input $X$, with $X_2\gg X_1$. 
The output $\Gamma_2$ is smaller now than that in Table~\ref{tab.211106.1}, though within errors they are again compatible. This clearly indicates the compatibility between the two methods for this $f_0(980)$ pole too. We also point out that the bare parameters  for $X=0.8$ in Table~\ref{tab.211107.1} are essentially undetermined, as indicated by the large intervals given between parenthesis, though the physical outputs  appear with an uncertainty of similar size as that in the other rows of the table.

\begin{table}[!htbp]
\begin{center}
\caption{Method F applied to the resonance $f_0(980)$ with the pole position in the RS \Rmnum{2} from Ref.~\cite{Guo:2012yt}, Eq.~\eqref{211107.1}. The  various entries are the same as in Table~\ref{tabflatt1}. The numbers between parenthesis indicate huge range of values  of the corresponding quantity  by varying $m_R$ and $\Gamma_R$ within errors.}
\label{tab.211107.1}
{ \begin{tabular}{|cc|ccccccc|}
\Xhline{1pt}
$X$&$\text{RS}$ &$\widetilde{\Gamma}_{\pi\pi}$(MeV) &$g_2$& $E_{f}$(MeV) &$\Gamma_{1}$(MeV) &$\Gamma_{2}$(MeV) &$X_{\pi\pi}$ &$X_{K\bar{K}}$\\
\Xhline{1pt}
\multirow{1}*{0.8}& $\text{\Rmnum{2}}$& $(430,3 \!\cdot\! 10^{6})$ &$(5.4,4  \!\cdot\! 10^4)$
&$(-160,-2 \!\cdot\! 10^{6})$& ~$71.6\pm18.2$ & $16.8\pm7.9$&~$0.028\pm0.007$&~$0.772\pm0.007$\\
\Xhline{1pt}
\multirow{1}*{0.6}& $\text{\Rmnum{2}}$&$218.4\pm51.9$ &\,$2.34\pm0.52$~&\!\!$-131.6\pm52.3$&~$68.8\pm16.9$&$12.8\pm7.0$&~$0.027\pm0.007$&~$0.573\pm0.007$\\
\Xhline{1pt}
\multirow{1}*{0.4}& $\text{\Rmnum{2}}$&$113.6\pm25.1$ &$0.83\pm0.13$ &$-52.6\pm20.7$&~$65.6\pm15.4$&~$9.6\pm5.1$&~$0.026\pm0.006$&~$0.374\pm0.006$\\
\Xhline{1pt}
\multirow{1}*{0.2}& $\text{\Rmnum{2}}$&~ $75.2\pm16.1$ &$0.27\pm0.03$ &$-24.2\pm11.8$&~$61.1\pm13.6$&~$5.1\pm2.6$&~$0.024\pm0.005$&~$0.176\pm0.005$\\
\Xhline{1pt}
\end{tabular}
\label{tab.211107.1}}
\end{center}
\end{table}
\subsection{The $a_0(980)$ resonance}
\label{sec.201009.2}

Let us explore the compositeness of the $a_0(980)$ similarly as done above in Sec.~\ref{sec.201009.1} for the $f_0(980)$.
The poles that we are going to consider next for the $a_0(980)$ stems from a recent coupled-channel partial-wave analysis of antiproton-proton annihilation data in Ref.~\cite{CrystalBarrel:2019zqh}, where the pole parameters and the partial-decay width of the $a_0(980)$ are discussed.
In the RS \Rmnum{2} Ref.~\cite{CrystalBarrel:2019zqh} reports the values
\begin{eqnarray}
  \label{211009.1}
m_{a_0}=1004.1\pm 6.67\,\, \text{MeV},~ \Gamma_{a_0}=97.2\pm6.01\,\,\text{MeV},~ \Gamma_{K\bar{K}}/\Gamma_{\pi\eta}=(13.8\pm3.5)\,\,\%,
\end{eqnarray}
while in the RS \Rmnum{3} the same reference provides
\begin{eqnarray}\label{eq:ratioa0}
m_{a_0}=1002.4\pm 6.55\,\, \text{MeV},~ \Gamma_{a_0}=127.0\pm7.08\,\,\text{MeV},~ \Gamma_{K\bar{K}}/\Gamma_{\pi\eta}=(14.9\pm3.9)\,\,\%,
\end{eqnarray}
with  the uncertainties added in quadrature.

\begin{table}[!htbp]
\begin{center}
\caption{{\small Method S applied for the resonance $a_0(980)$ with pole positions  from Ref.~\cite{CrystalBarrel:2019zqh}, Eqs.~\eqref{211009.1} and \eqref{eq:ratioa0}, in different RSs (column 2): We take input values for $X$ (column 1), and predict $|\ga_i|$ (columns 3, 4),  $\Gamma_i$ (columns 5, 6), and $X_i$ (columns 7, 8).} }
\begin{tabular}{|cc|cccccc|}
\Xhline{1pt}
~~$X~~~$&~~~$\text{RS}$&~~~$|\gamma_{\pi\eta}|(\text{GeV})$~~~&$|\gamma_{K\bar{K}}|(\text{GeV})$~~~&$\Gamma_{1}(\text{MeV})$&~~~$\Gamma_{2}(\text{MeV})$&~~~$X_{\pi\eta}$&~~~$X_{K\bar{K}}$~~~\\
\Xhline{1pt}
\multirow{2}*{1.0}&~~~$\text{\Rmnum{2}}$&~~~$3.78\pm0.08$~~~&$5.83\pm0.05$~~~&$189.3\pm8.2$&~~~$92.1\pm6.3$&~~~$0.169\pm0.006$&~~~$0.831\pm0.006$~~~\\
&~~~$\text{\Rmnum{3}}$&~~~$2.01\pm0.12$~~~&$5.11\pm0.07$~~~&$53.9\pm6.4$&~~~$73.1\pm5.7$&~~~$0.049\pm0.006$&~~~$0.951\pm0.006$~~~\\
\Xhline{1pt}
\multirow{2}*{0.8}&~~~$\text{\Rmnum{2}}$&~~~$3.57\pm0.07$~~~&$5.15\pm0.04$~~~&$169.1\pm7.0$&~~~$71.9\pm4.9$&~~~$0.151\pm0.005$&~~~$0.649\pm0.005$~~~\\
&~~~$\text{\Rmnum{3}}$&~~~$2.30\pm0.09$~~~&$4.50\pm0.06$~~~&$70.4\pm5.5$&~~~$56.6\pm4.4$&~~~$0.064\pm0.006$&~~~$0.736\pm0.006$~~~\\
\Xhline{1pt}
\multirow{2}*{0.6}&~~~$\text{\Rmnum{2}}$&~~~$3.35\pm0.06$~~~&$4.37\pm0.03$~~~&$149.0\pm5.8$&~~~$51.8\pm5.8$&~~~$0.133\pm0.005$&~~~$0.467\pm0.005$~~~\\
&~~~$\text{\Rmnum{3}}$&~~~$2.56\pm0.07$~~~&$3.78\pm0.05$~~~&$86.9\pm4.9$&~~~$40.1\pm3.1$&~~~$0.079\pm0.005$&~~~$0.521\pm0.005$~~~\\
\Xhline{1pt}
\multirow{2}*{0.4}&~~~$\text{\Rmnum{2}}$&~~~$3.12\pm0.05$~~~&$3.41\pm0.02$~~~&$128.8\pm4.6$&~~~$31.6\pm2.2$&~~~$0.115\pm0.004$&~~~$0.285\pm0.004$~~~\\
&~~~$\text{\Rmnum{3}}$&~~~$2.79\pm0.06$~~~&$2.90\pm0.04$~~~&$103.5\pm4.5$&~~~$23.5\pm1.9$&~~~$0.094\pm0.005$&~~~$0.306\pm0.005$~~~\\
\Xhline{1pt}
\multirow{2}*{0.2}&~~~$\text{\Rmnum{2}}$&~~~$2.86\pm0.05$~~~&$2.05\pm0.01$~~~&$108.6\pm3.7$&~~~$11.4\pm0.8$&~~~$0.097\pm0.003$&~~~$0.103\pm0.003$~~~\\
&~~~$\text{\Rmnum{3}}$&~~~$3.01\pm0.06$~~~&$1.58\pm0.04$~~~&$120.0\pm4.5$&~~~$7.0\pm0.7$&~~~$0.109\pm0.005$&~~~$0.091\pm0.005$~~~\\
\Xhline{1pt}
\end{tabular}
\label{tabff4}
\end{center}
\end{table}

\begin{table}[!htbp]
\begin{center}
\caption{ Method F applied to the resonance $a_0(980)$  with pole positions from Ref.~\cite{CrystalBarrel:2019zqh}, Eqs.~\eqref{211009.1} and \eqref{eq:ratioa0}, in different RSs (column 2): We calculate the bare width $\widetilde{\Gamma}_1$ (column 3), the bare coupling squared $g_2$ (column 4),  $E_{f}$ (column 5), $\Gamma_1$ (column 6),  $\Gamma_2$ (column 7), $X_1$ (column 8) and $X_{2}$ (column 9). The value of $X$ taken as input is given in the first column. The dashes in the second row indicate that no solution is found for $X=1$ when the pole lies in the RS \Rmnum{2}.}
\begin{tabular}{|cc|ccccccc|}
\Xhline{1pt}
$X~$&$\text{RS}$&$\widetilde{\Gamma}_{\pi\eta}$(MeV)&~$g_2$&$E_{f}(\text{MeV})$&$\Gamma_{1}$(MeV)&$\Gamma_{2}$(MeV)&$X_{\pi\eta}$&$X_{K\bar{K}}$\\
\Xhline{1pt}
\multirow{2}*{1.0}&~$\text{\Rmnum{2}}$&~$--$~&$--$~&$--$&~$--$&~$--$&~$--$&~$--$\\
&~$\text{\Rmnum{3}}$&~$14.3\pm7.8$~&$0.83\pm0.04$~&$58.2\pm4.5$&~$18.9\pm10.5$&~$108.4\pm10.9$&~$0.017\pm0.010$&~$0.983\pm0.010$\\
\Xhline{1pt}
\multirow{2}*{0.8}&~$\text{\Rmnum{2}}$&~$740.5\pm42.2$~&$5.16\pm0.51$~&$-237.1\pm39.4$&~$157.2\pm8.8$&~$60.0\pm7.4$&~$0.141\pm0.007$&~$0.659\pm0.007$\\
&~$\text{\Rmnum{3}}$&~$35.9\pm7.3$~&$0.66\pm0.03$~&$49.3\pm4.6$&~$45.6\pm9.7$&~$81.4\pm9.3$&~$0.042\pm0.009$&~$0.758\pm0.009$\\
\Xhline{1pt}
\multirow{2}*{0.6}&~$\text{\Rmnum{2}}$&~$289.7\pm8.5$~&$1.54\pm0.06$~&$-61.6\pm10.0$&~$146.2\pm7.2$&~$49.0\pm5.6$&$0.131\pm0.005$&~$0.469\pm0.005$\\
&~$\text{\Rmnum{3}}$&~$59.9\pm5.6$~&$0.49\pm0.02$~&$39.2\pm4.3$&~$72.7\pm7.1$&~$54.3\pm6.2$&~$0.066\pm0.007$&~$0.534\pm0.007$\\
\Xhline{1pt}
\multirow{2}*{0.4}&~$\text{\Rmnum{2}}$&~$177.4\pm5.6$~&$0.64\pm0.01$~&$-18.2\pm5.8$&~$129.9\pm5.4$&~$32.7\pm3.5$&$0.116\pm0.004$&~$0.284\pm0.004$\\
&~$\text{\Rmnum{3}}$&~~~$85.2\pm4.7$~&$0.30\pm0.01$~&$28.6\pm4.1$&~$96.9\pm5.3$&~$30.1\pm3.4$&~$0.089\pm0.006$&~$0.311\pm0.006$\\
\Xhline{1pt}
\multirow{2}*{0.2}&~$\text{\Rmnum{2}}$&~$120.2\pm3.8$~&$0.18\pm0.002$~&$4.0\pm4.3$&~$109.9\pm3.8$&~$12.7\pm1.4$&~$0.099\pm0.003$&~$0.101\pm0.003$\\
&~$\text{\Rmnum{3}}$&~$113.6\pm4.6$~&$0.10\pm0.01$~&$16.7\pm4.0$&~$118.6\pm4.6$&~$8.4\pm1.1$&~$0.108\pm0.005$&~$0.092\pm0.005$\\
\Xhline{1pt}
\end{tabular}
\label{tabflatt3}
\end{center}
\end{table}

Let us denote by 1 the lighter channel $\pi\eta$, and by 2 the heavier one $K\bar{K}$.
Combining Eq.~(\ref{equ:square2}) for the total compositeness and Eq.~(\ref{equ:square1}) for the full width, given input values for the former between 0.2 and 1.0 in steps of 0.2, 
we derive a series of partial compositeness coefficients and decay widths for the $a_{0}(980)$ in Table \ref{tabff4} by applying the method S.\footnote{The same remark as for the $f_0(980)$ in the footnote \ref{foot.211008.1} can also be applied for the $a_0(980)$ when its pole is taken in the RS \Rmnum{2} for applying Eq.~\eqref{equ1} to calculate $X_2$.
  The point is that the $a_0(980)$ mass in Eq.~\eqref{211009.1} is clearly larger than the $K\bar{K}$ threshold but the width of the resonance is much larger than the difference between $m_{a_0}$ and the $K\bar{K}$ threshold.}
We find that  $X_2$ is rather similar for the RS \Rmnum{2} and \Rmnum{3} calculations. We also obtain that $X_2\gg X_1$, except for $X=0.2$ in which case they are quite close to each other. 
This tells us that the $K\bar{K}$ component typically dominates over the $\pi\eta$ one for the resonance $a_0(980)$.

The results corresponding to the method F for the $a_0(980)$ case are organized in Table~\ref{tabflatt3}.
One then finds that the values of $X_2$ obtained in Tables~\ref{tabflatt3} and \ref{tabff4} are very close independently of the RS
in which the poles lie.
 It turns out that $X_2$ is typically much larger than $X_1$  except when $X$ becomes small, like $X=0.2$ in Table~\ref{tabflatt3}, and then both $X_i$ are rather close to each other. 
 Regarding the $\Gamma_i$ it is clear that for the pole in the RS \Rmnum{2} the values obtained now in Table~\ref{tabflatt3} are remarkably similar to those in Table~\ref{tabff4}. However, they clearly differ for $X\geq 0.8$ for the RS \Rmnum{3} case, with no solution found  for $X=1$. Of course, $X=0$ cannot be reproduced in any method S or F due to the finite width of the $a_0(980)$, which requires non-vanishing couplings.  

As indicated above one should distinguish between the bare and dressed parameters in a Flatt\'e parametrization.
    This reflects in the fact that the bare width $\widetilde{\Gamma}_1$ can be much larger than the total width $\Gamma_R$.
    This is particularly true when the pole position is set to lie in the RS \Rmnum{2}, as it is clear from the Tables~\ref{tabflatt1}, \ref{tab.211107.1} and \ref{tabflatt3}.
    Here in addition one has to properly relate $\Gamma_1$ and $\Gamma_{\pi\pi}$, cf. Eq.~\eqref{211110.1}.
    Nonetheless, these important points have been overlooked in the literature.

Before finishing this section we should mention that in some studies the $a_0(980)$ lies in the RS \Rmnum{4}, so that there is no an interval along the physical real $s$-axis within the radius of convergence of the Laurent series around the $a_0(980)$.
In such a scenario it is not justified to apply Eq.~\eqref{equ1} for the calculation of $X_2$, as discussed in detail in Ref.~\cite{Guo:2015daa}, and it is not either clear how to connect the $\Gamma_i$ with the physical partial-decay widths. Then, it follows that our method 
(as well as the older Ref.~\cite{Baru:2003qq})
cannot  be applied and the conclusions  thereof do not hold.
In the studies based on unitarization of SU(3) and U(3) chiral perturbation theory it is observed that the pole for the $a_0(980)$ appears in the 2nd RS if only tree level amplitudes are kept.
This is necessarily the case when unitarizing the leading order amplitudes \cite{Oller:1998zr,Oller:1997ti,Oller:1999ag,Guo:2016zep}.
However, once loop contributions are accounted for the pole moves to the RS \Rmnum{4} \cite{Guo:2011pa,Guo:2012yt,Guo:2016zep}.
This is also the case in the $K$-matrix analysis of Ref.~\cite{Dudek:2016cru}.
The more recent studies among those quoted here \cite{Guo:2016zep,Dudek:2016cru} also reproduce the energy levels from lattice QCD simulations.
When the $a_0(980)$ lies in the RS \Rmnum{4} it manifests on the physical real $s$-axis as a strong cusp effect.

\section{Results and discussions using the branching ratio $\rx$ as input}
\label{sec.211126.1}

Here, we directly explore the more stringent scenario of taking $r_{\rm exp}$, together with the usual reproduction of the pole position $E_R$, in order to fix all the parameters both in the S and F methods.
The direct consideration of $\rx$ as input is  an interesting check on the consistency of interpreting the $\Gamma_i$, calculated from the pole parameters, in connection with the experimental $\rx$  for a pole lying in the RS \Rmnum{2}.
As shown below, the resulting physical-partial-decay widths have values within meaningful ranges, such that they are positive
and smaller than the total width, while the $X_i$ and their sum $X$ also lie within the allowed interval $[0,1]$.\footnote{Otherwise, if not correctly understanding $\rx$ as proposed for a RS \Rmnum{2} pole, there will be no solution within the method F developed in Sec.~\ref{sec.201009.3}.}
This situation is similar to the one obtained before when taking $X$ as input,
 where all the Flatt\'e parameters are fixed without the need to connect the $\Gamma_i$ with $\rx$, which comes out as an output.

 We also show below that the procedure of providing $\rx$ as input is not completely satisfactory because the output value for the renormalized coupling squared to $K\bar{K}$, $|\gamma_2|^2$, can be very sensitive to small variations in the input. As a result, the calculation of $X_2$ by using Eq.~\eqref{equ1} does not come out very accurate, particularly  for the $f_0(980)$. We can circumvent this limitation by then using the method of the spectral function $\omega(E)$ to estimate $X$ with better precision. With this choice we also update values for the resonance inputs compared with the 
 almost two decades old  Ref.~\cite{Baru:2003qq}, where the spectral density function to calculate the compositeness was applied for the first time to the $f_0(980)$ and $a_0(980)$ resonances. 
 In this respect, it is important to indicate that nowadays we have at our disposal the precise determination of the $f_0(980)$ pole from the Roy-like GKPY equations, Eq.~\eqref{equ22}. 

\subsection{The $f_0(980)$ resonance}
\label{sec.211126.2}

Now, let us directly combine the knowledge of $E_R$ with the input for the branching ratio
$r_{\rm exp}$,  $\rx=0.52\pm 0.12$ \cite{Aubert:2006nu} and $\rx=0.75^{+0.11}_{-0.13}$ \cite{BES:2005iaq}, as well as $\rx=0.68$ from the theoretical analysis of Ref.~\cite{Oller:1997ti}. 
Notice that in this way we consider a broad range of values for $\rx$ from around 0.40 up to 0.86, taking into account the spread in the central values within one standard deviation. As a result, instead of selecting a specific experimental input,
rather we  try to extract an image for the nature of the $f_0(980)$ compatible with the present rather imprecise  experimental knowledge in $\rx$.  

We start by applying the method S.
It is clear that given $r_{\rm exp}$ and $\Gamma_{R}$ one can solve for the couplings $|\gamma_i|$, $i=1,\,2$, and calculate the partial compositeness coefficients $X_i$.
 Since the $f_0(980)$ poles that we are considering lie in the RS \Rmnum{2} we have that $\Gamma_{\pi\pi}=(2-\rx)\Gamma_{R}$ and $\Gamma_{K\bar{K}}=(1-\rx)\Gamma_{R}$, and similarly for the RS \Rmnum{2} pole for the $a_0(980)$, $\Gamma_{\pi\eta}=(2-\rx)\Gamma_{R}$. For the $a_0(980)$ pole in the RS \Rmnum{3} we have instead  $\Gamma_{\pi\eta}=\rx\Gamma_{R}$. By combining Eqs. (\ref{equ1}), (\ref{equ33}) and (\ref{equ:square})
 the value of the compositeness coefficients $X_{1}$ and $X_{2}$ can be written as
\begin{eqnarray}
X_{1}&=&\frac{8\pi (2-\rx)\Gamma_{R} m_R^2}{p_1(m_R^2)} \left|\frac{\partial G_{1}(s)}{\partial s}\right|_{s=s_{R}},~\text{RS \Rmnum {2}}~,\\
X_{1}&=&\frac{8\pi \rx\Gamma_{R} m_R^2}{p_1(m_R^2)} \left|\frac{\partial G_{1}(s)}{\partial s}\right|_{s=s_{R}},~\text{RS \Rmnum {3}}~,\nn\\
X_{2}&=&\frac{16\pi^2 (1-\rx)\Gamma_{R}}{\int_{m_{1}+m_{2}}^{m_{R}+2\Gamma_{R} }dW\frac{p(W^{2})/W^{2}}{(m_{R}-W)^{2}+\Gamma_{R}^2/4}}
\left|\frac{\partial G_{2}(s)}{\partial s}\right|_{s=s_{R}}.
\end{eqnarray}
The results are organized in Table~\ref{tabff3} with $m_R$ and $\Gamma_R$  taken from Eq.~\eqref{equ22}, corresponding to the determination in Ref.~\cite{GarciaMartin:2011jx}. For this and the rest of tables in this section the error bars given to $X_2$ and $X$ contain also the propagation of the error in the determination of $\rx$ from Refs.~\cite{Aubert:2006nu,BES:2005iaq}.
It follows from this table that  $X_2$ is affected by large error bars, around a 50\%, so that the resulting $X$ is not pined down accurately and the calculation is rather indicative.  Only for $\rx=0.52\pm 0.12$ \cite{Aubert:2006nu} one has that $X$ can be larger than 0.5 within errors, with $X=0.49\pm 0.20$. For the other experimental value $\rx=0.75^{+0.11}_{-0.13}$ \cite{BES:2005iaq} the compositeness $X=0.27\pm 0.11$ and it is smaller than 0.5.

\begin{center}
\begin{figure}[!htbp]
	\centering
\caption{ The total compositeness $X$ is plotted as a function of the input  branching ratio ($r_\text{exp})$ when applying the method S. Panel (a) corresponds to the $f_0(980)$: The dashed line is for the pole from Ref.~\cite{GarciaMartin:2011jx}, Eq.~\eqref{equ22}, and the dashed one is for the pole of Ref.~\cite{Guo:2012yt}, Eq.~\eqref{211107.1}.  Panel (b) corresponds to the $a_0(980)$: The poles for the $a_0(980)$ are given in Eqs.~\eqref{211009.1} and \eqref{eq:ratioa0} for the RSs \Rmnum{2} and \Rmnum{3}, respectively, and they are taken from Ref.~\cite{CrystalBarrel:2019zqh}.} \label{pic_X_r}
	\begin{minipage}[t]{1\linewidth}
		\centering
		\subfigure[]{
			\includegraphics[width=0.45\textwidth]{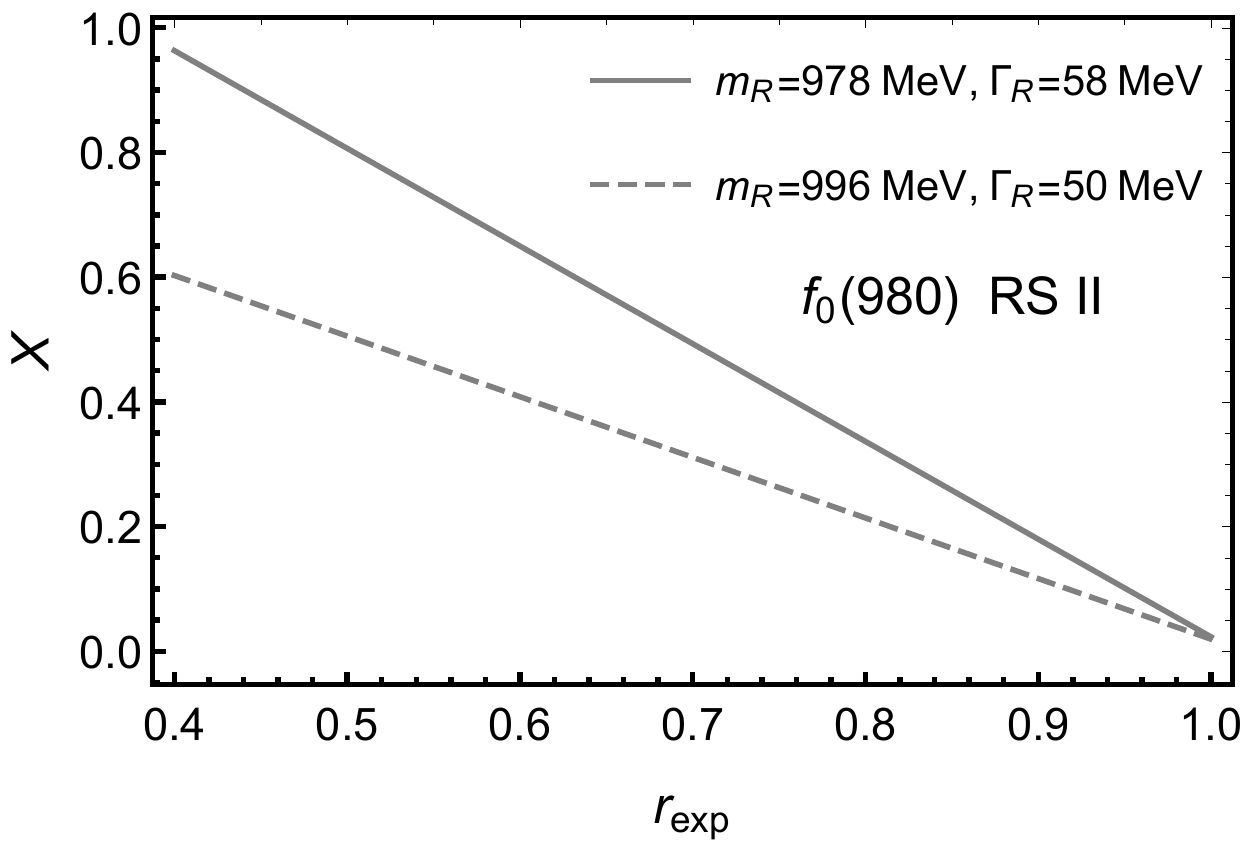}}
		\hspace{0.45in}
		\subfigure[]{
			\includegraphics[width=0.45\textwidth]{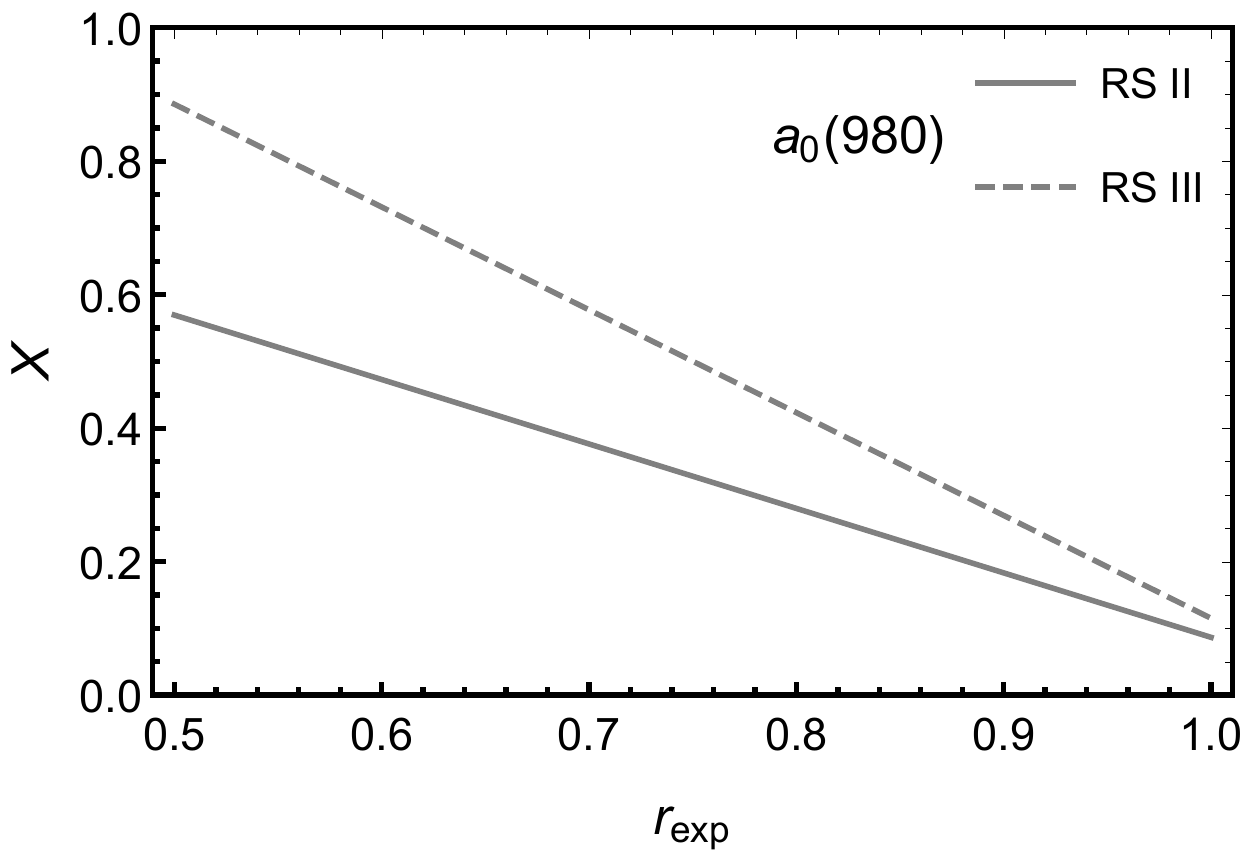}}
	\end{minipage}
\end{figure}
\end{center}

We show in the left panel of Fig.~\ref{pic_X_r} the total compositeness $X$ as a function of $r_{\rm exp}$ for the $f_0(980)$ with the method S, and employing the central values for $m_R$ and $\Gamma_R$ for the different poles considered.  Thus, the requirement that $r_{\rm exp}>0.4$ implies that $X\lesssim 0.6$ for the pole in Eq.~\eqref{equ22} in the RS \Rmnum{2}.
It is also clear from Fig.~\ref{pic_X_r} the linear decrease of $X$ with $r_{\rm exp}$.
This can be easily understood by noticing that both the
partial-decay widths and partial compositeness coefficients are proportional to $|\gamma_i|^2$, which can be written in turn as $\Gamma_i/\theta_i$. For given values for the mass and width of the resonance the $\theta_i$ is just a measure of the available phase space for the decay to the channel $i$, and then $\theta_2\ll \theta_1$. In this way, for a pole in the RS \Rmnum{3},
\begin{align}
\label{211103.1}
X&=\frac{\Gamma_R r_{\rm exp}}{\theta_1}\left|\frac{\partial G_1}{\partial s}\right|_{s=s_R}
+\frac{\Gamma_R (1-r_{\rm exp})}{\theta_2}\left|\frac{\partial G_2}{\partial s}\right|_{s=s_R}\\
&=\frac{\Gamma_R }{\theta_2}\left|\frac{\partial G_2}{\partial s}\right|_{s=s_R}
-r_{\rm exp}\Gamma_R\left(\frac{1}{\theta_2}\left|\frac{\partial G_2}{\partial s}\right|_{s=s_R}
-\frac{1}{\theta_1}\left|\frac{\partial G_1}{\partial s}\right|_{s=s_R}\right)~.\nn
\end{align}
This is a linear dependence with $r_{\rm exp}$, and the coefficient multiplying the latter is negative because  both $1/\theta_2$ and $\left|\partial G_2/\partial s\right|_{s_R}$ are basically proportional to $|s_R/4-m_K^2|^{-1/2}$, cf. Eq.~\eqref{211106.1}.\footnote{Equation~\eqref{211103.1} was written for a pole in the RS \Rmnum{3}.
  For a pole lying in the RS \Rmnum{2} one has to take into account that $\Gamma_1=\Gamma_R(2-\rx)$ and, then,  instead of Eq.~\eqref{211103.1} we have
  \begin{align}
\label{211122.1}
    X&=2\frac{\Gamma_R }{\theta_1}\left|\frac{\partial G_1}{\partial s}\right|_{s=s_R}
    +\frac{\Gamma_R }{\theta_2}\left|\frac{\partial G_2}{\partial s}\right|_{s=s_R}
-r_{\rm exp}\Gamma_R\left(\frac{1}{\theta_2}\left|\frac{\partial G_2}{\partial s}\right|_{s=s_R}
+\frac{1}{\theta_1}\left|\frac{\partial G_1}{\partial s}\right|_{s=s_R}\right)~,
  \end{align}
 so that the linear decrease of $X$ with $\rx$ follows.
}


\begin{table}[!htbp]
\begin{center}
\caption{ Method S applied to the resonance $f_0(980)$ with pole position in the RS \Rmnum{2} (column 2) from Ref.~\cite{GarciaMartin:2011jx}, Eq.~\eqref{equ22}: By reproducing values given for $\rx$ (column 1), we calculate $|\gamma_i|$ (columns 3, 4), $\Gamma_{i}$ (columns 5, 6), and  $X_{i}$ (columns 7, 8).}
\begin{tabular}{|cc|cccccc|}
	\Xhline{1pt}
~~$r_{\rm
exp}$&~~~$\text{RS}$&~~~$|\gamma_{1}|(\text{GeV})$~~~&$|\gamma_{2}|(\text{GeV})$~~~&$\Gamma_{1}(\text{MeV})$&~~~$\Gamma_{2}(\text{MeV})$&~~~$X_{\pi\pi}$&~~~$X_{K\bar{K}}$~~~\\
	\Xhline{1pt}
\multirow{1}*{0.52~\cite{Aubert:2006nu}}&~~~$\text{\Rmnum{2}}$&~~~$2.03\pm0.18$~~~&$3.62\pm0.32$~~~&$79.9\pm14.5$&~~~$25.9\pm4.7$&~~~$0.031\pm0.006$&~~~$0.46 \pm 0.20$~~~\\
\Xhline{1pt}	\multirow{1}*{0.68~\cite{Oller:1997ti}}&~~~$\text{\Rmnum{2}}$&~~~$1.92\pm0.17$~~~&$2.95\pm0.26$~~~&$71.3\pm13.0$&~~~$17.3\pm3.1$&~~~$0.028\pm0.005$&~~~$0.31\pm0.13 $~~~\\
\Xhline{1pt}
\multirow{1}*{0.75~\cite{BES:2005iaq}}&~~~$\text{\Rmnum{2}}$&~~~$1.87\pm0.17$~~~&$2.61\pm0.23$~~~&$67.5\pm12.3$&~~~$13.5\pm2.5$&~~~$0.026\pm0.005$&~~~$0.24 \pm 0.11$~~~\\
	\Xhline{1pt}
\end{tabular}
\label{tabff3}
\end{center}
\end{table}

\begin{table}[!htbp]
\begin{center}
\caption{ Method F applied to the resonance $f_0(980)$ with the pole position in the RS \Rmnum{2} from Ref.~\cite{GarciaMartin:2011jx}, Eq.~\eqref{equ22}: The branching ratio $r_{\text{exp}}$ is taken as input (column 1).
  We then calculate the bare partial-decay width $\widetilde{\Gamma}_1$ (column 2), the bare coupling squared $g_2$ (column 3), $E_f$ (column 4),  the renormalized $\Gamma_1$ (column 5) and $\Gamma_2$ (column 6), and the partial compositeness coefficients $X_1$ (column 7) and  $X_2$ (column 8). The numbers between parenthesis indicate large ranges of the corresponding quantities by varying $m_R$ and $\Gamma_R$ within errors.  }
\begin{tabular*}{\linewidth}{@{\extracolsep{\fill}}|c|ccccccc|}  
\Xhline{1pt}
$r_{\text{exp}}$ &$\widetilde{\Gamma}_{\pi\pi}\,(\text{MeV})$&$g_2$ &$E_{f}\,(\text{MeV})$& $\Gamma_{1}\,(\text{MeV})$ &$\Gamma_{2}\, (\text{MeV})$ &$X_{\pi\pi}$ &$X_{K\bar{K}}$\\
\Xhline{1pt}
\multirow{1}*{0.52 \cite{Aubert:2006nu}} &$(71, 2623)$ 
&$(0.36, 26.8)$ 
&$(-21.2, 1167.2)$ &$79.9\pm14.5$ &$25.9\pm4.7$ &$0.030\pm0.006$ &$0.48\pm 0.22$\\
\Xhline{1pt}
\multirow{1}*{0.68 \cite{Oller:1997ti}} &$113.8\pm38.3$ &$0.69\pm0.45$ &$(-66.7, 6.3)$ &$71.3\pm13.0$ &$17.3\pm3.1$ &$0.028\pm0.005$ &$0.35\pm0.18$\\
\Xhline{1pt}
\multirow{1}*{0.75 \cite{BES:2005iaq}} &$89.4\pm20.4$ &$0.41\pm0.17$ &$(-42.9, 7.7)$ &$67.5\pm12.3$ &$13.5\pm2.5$ &$0.026\pm0.005$ &$0.26\pm0.14$\\
\Xhline{1pt}
\end{tabular*}
\label{tab.211111.1}
\end{center}
\end{table}

Now we apply the method F with $\rx$ as input to the same $f_0(980)$ pole given in Eq.~\eqref{equ22}, and the  results are presented in the Table~\ref{tab.211111.1}. In the columns 2--4 we provide the resulting parameters characterizing the Flatt\'e parametrization, and in the rest of columns we calculate the outputs in the form of the partial-decay widths $\Gamma_i$ and partial compositeness coefficients $X_i$. We notice that for the smallest
$\rx=0.52$ \cite{Aubert:2006nu} the bare parameters are much more poorly determined than those in the other cases. Nonetheless, the outputs have uncertainties of similar sizes as for the other values taken for $\rx$. The resulting values for $X_2$ in Tables~\ref{tab.211111.1} and \ref{tabff3} turn out to be remarkably close to each other, clearly showing the compatibility between the methods F and S.

Let us compare Table~\ref{tab.211111.1} with Table~\ref{tabflatt1}, where $X$ is taken as input and no use of the interpretation of the $\Gamma_i$ for a pole in the RS \Rmnum{2}, according to Eq.~\eqref{211110.1}, is done to fix the Flatt\'e parametrization.
One can then observe that similar  values of $\Gamma_{K\bar{K}}$, which fixes $\rx=1-\Gamma_{K\bar{K}}/\Gamma_R$, lead to close values of $X_2$ in both tables. This agreement is of course an indication that the equations are correctly solved,
because the same results are obtained by taking corresponding values of $X$ or $\rx$.

\begin{table}[!htbp]
\begin{center}
	\caption{{\small Method S applied to the resonance $f_0(980)$ with pole position in the RS \Rmnum{2}  from Ref.~\cite{Guo:2012yt}, Eq.~\eqref{211107.1}. The various entries are the same as in Table~\ref{tabff3}.} \label{tab.211106.2}}
\begin{tabular}{|cc|cccccc|}
	\Xhline{1pt}
	~~$r_{\rm exp}$&~~~$\text{RS}$&~~~$|\gamma_{\pi\pi}|(\text{GeV})$~~~&$|\gamma_{K\bar{K}}|(\text{GeV})$~~~&$\Gamma_{1}(\text{MeV})$&~~~$\Gamma_{2}(\text{MeV})$&~~~$X_{\pi\pi}$&~~~$X_{K\bar{K}}$~~~\\
	\Xhline{1pt}
\multirow{1}*{0.52~\cite{Aubert:2006nu}}&~~~$\text{\Rmnum{2}}$&~~~$2.05\pm0.22$~~~&$4.71\pm0.70$~~~&$82.9\pm18.2$&~~~$26.9\pm5.9$&~~~$0.033\pm0.007$&~~$0.72\pm 0.34 $~~~\\
\Xhline{1pt}	\multirow{1}*{0.68~\cite{Oller:1997ti}}&~~~$\text{\Rmnum{2}}$&~~~$1.94\pm0.21$~~~&$3.85\pm0.58$~~~&$73.9\pm16.2$&~~~$17.9\pm3.9$&~~~$0.029\pm0.006$&~~~$0.48 \pm0.23 $~~~\\
\Xhline{1pt}
\multirow{1}*{0.75~\cite{BES:2005iaq}}&~~~$\text{\Rmnum{2}}$&~~~$1.88\pm0.21$~~~&$3.40\pm0.51$~~~&$70.0\pm15.3$&~~~$14.0\pm3.1$&~~~$0.027\pm0.006$&~~~$0.38 \pm0.18 $~~~\\
	\Xhline{1pt}
\end{tabular}
\end{center}
\end{table}

\begin{table}[!htbp]
\begin{center}
\caption{{\small Method F applied to the resonance $f_0(980)$ with the pole position in the RS \Rmnum{2} from Ref.~\cite{Guo:2012yt}, Eq.~\eqref{211107.1}: The entries are the same as those in Table~\ref{tab.211111.1}, except for the values of $\rx$ in the first column because  solutions  are found only for $\rx>0.82$. The values of $\rx$ taken do not come from any experiment and are considered due to theoretical reasons.}
  \label{tabflatt2}}
\begin{tabular*}{\linewidth}{@{\extracolsep{\fill}}|c|ccccccc|}  
\Xhline{1pt}
$r_{\text{exp}}$& $\widetilde{\Gamma}_{\pi\pi} (\text{MeV})$ &$g_2$& $E_{f} (\text{MeV})$ & $\Gamma_{1} (\text{MeV})$ &$\Gamma_{2} (\text{MeV})$ &$X{\pi\pi}$ &$X_{K\bar{K}}$\\
\Xhline{1pt}
\multirow{1}*{0.86} &$(45.8, 2632)$ &$(0.13, 35.6)$ &$(-2358, 0.3)$ &$63.8\pm14.0$ &$7.8\pm1.7$ &$0.026\pm0.005$ &$0.329\pm0.140$\\
\Xhline{1pt}
\multirow{1}*{0.91} &$(41.9, 407.6)$ &$(0.08, 6.64)$ &$(-429.4, 1.3)$ &$61.0\pm13.4$ &$5.1\pm1.1$ &$0.025\pm0.005$ &$0.239\pm0.123$\\
\Xhline{1pt}
\multirow{1}*{0.96} &$66.1\pm13.0$ &$0.16\pm0.12$ &$-20.8\pm14.4$ &$58.2\pm12.8$ &$2.3\pm0.5$ &$0.023\pm0.005$ &$0.105\pm0.058$\\
\Xhline{1pt}
\end{tabular*}
\end{center}
\end{table}

 Let us move on and consider the $f_0(980)$ pole position from Ref.~\cite{Guo:2012yt} applying first the method S, with the results given in Table~\ref{tab.211106.2}.
 We already noticed when comparing Tables~\ref{tabff1} and \ref{tab.211106.1} that the $\Gamma_i$'s were larger for the former despite the width of the pole in Eq.~\eqref{equ22} is smaller.
 By comparing Tables~\ref{tabff3} and \ref{tab.211106.2} with $\rx$ as input we observe that the partial-decay widths $\Gamma_i$'s have close values between them and that the differences translates into the central values  of $X_2$, which are  always larger in Table~\ref{tab.211106.2} than the ones in Table~\ref{tabff3}. Thus, a somewhat more prominent role of the $K\bar{K}$ component in the case of the $f_0(980)$ pole  in Eq.~\eqref{211107.1} arises.

When applying the method F to the $f_0(980)$ pole from Ref.~\cite{Guo:2012yt}, Eq.~\eqref{211107.1}, there are no acceptable solutions for the input values of $\rx$ from Refs.~\cite{Aubert:2006nu,BES:2005iaq,Oller:1997ti}, with solutions found only  for larger values of $\rx$.
 Indeed, we see from Eq.~\eqref{211111.1} that the central value for $\rx$ calculated from the information given in Ref.~\cite{Guo:2012yt} is $0.91$, though the error estimated is too large to extract more stringent quantitative conclusions.   Our results are given in the Table~\ref{tabflatt2}, where $X_2$ varies by around a factor of 3 when changing $\rx$ by less than a 10\%, from $\rx=0.96$ to $0.86$.  For similar values of $\Gamma_{K\bar{K}}$ the results now and those in Table~\ref{tab.211107.1} agree well within errors, and give rise to rather small values for $X_2$.

Interestingly, we can check our procedure with the full characterization of the $f_0(980)$ pole given in Ref.~\cite{Guo:2012yt}, in which the residues, in addition to the pole position, are given. In this reference one has $|\gamma_1|=1.8^{+0.2}_{-0.3}$~GeV and $|\gamma_2/\gamma_1|=2.6^{+0.2}_{-0.3}$.
Precisely, the input value for $\rx=2-\frac{\Gamma_1}{\Gamma_R}=0.909$, cf. Eq.~\eqref{211111.1}, is equivalent to providing the central value of $|\gamma_1|$ as input, and solving the equation
\begin{align}
\label{211127.1}
|\gamma_1|^2&=\beta g_1=\beta \frac{\widetilde{\Gamma}_1 8\pi m_R^2}{p_1(m_R)}~,
\end{align}
instead of Eq.~\eqref{211110.2}.
The result is the same as the one given in the third row of Table~\ref{tabflatt2} corresponding  to a value $|\gamma_2|=\sqrt{32 \pi m_K^2 \beta g_2}=2.77$~GeV  [which straightforwardly translates into the central value for $X_2$ by using Eq.~\eqref{equ1}]. We notice that this value is a factor of 1.7 smaller  than the reported one in Ref.~\cite{Guo:2012yt}. Its square is therefore a factor of 2.84 smaller than the value $|\gamma_2|^2=(4.68$~GeV)$^2$ actually obtained in Ref.~\cite{Guo:2012yt} by solving for the residues of the partial-wave amplitudes.

However, instead of $|\gamma_1|$ we can  take $|\gamma_2|$ as input and solve the equation
\begin{align}
\label{211127.2}
|\gamma_2|^2&=32\pi m_K^2\beta g_2~,
\end{align}
instead of Eq.~\eqref{211127.1}.
Doing this exercise with the central value  $|\gamma_2|=4.68$~GeV \cite{Guo:2012yt}, corresponding to a much larger $X_2=0.68$, we find that $|\gamma_1|=1.875$~GeV, which is indeed perfectly compatible with the value $|\gamma_1|=1.8^{+0.2}_{-0.3}$~GeV given in Ref.~\cite{Guo:2012yt}. The branching ratio $\rx$ obtained with this value of $|\gamma_1|$ is 0.81, instead of 0.91 for $|\gamma_1|=1.8$~GeV (third row in Table~\ref{tabflatt2}).\footnote{At first sight it could seem strange that $|\gamma_1|=1.875>1.8$~GeV has $\rx=0.81<0.91$, respectively. This is because for $|\gamma_1|=1.875$~GeV the coupling $|\gamma_2|=4.68>2.77$~GeV and the decay width to $K\bar{K}$, $\Gamma_1-\Gamma_R$, has increased. 
  The fact that the width to $K\bar{K}$ has not increased by a relative factor of 2.84 as $|\gamma_2|^2$ does is a reflection of some divergence  between the F and S methods for large $K\bar{K}$ couplings in the case of the pole in Eq.~\eqref{211107.1}.
  This point  is also clear from the absence of solutions for $\rx=0.52$, 0.68 and $0.75$ within the method F, although they are found for the method S in Table~\ref{tab.211106.2}. Some divergence in the results when taking $X$ as input can also be seen for larger values of $X\gtrsim 0.6$ (and hence of $|\gamma_2|$) by comparing the central values for the $\Gamma_i$ between Tables~\ref{tab.211106.1} and \ref{tab.211107.1}. }

After  this check we reconsider the application of the method F to the $f_0(980)$ pole in Eq.~\eqref{equ22} and discuss the central values of $|\gamma_1|\,(|\gamma_2|)$ for the input values of $\rx$ in Table~\ref{tab.211111.1}. For the extreme values $\rx=0.52$,  and $0.75$ we find $1.96\,(3.88)$~GeV, and $1.78\,(2.40)$~GeV, respectively.
With a small variation of a 5\% in $\rx$, between the values $0.95 \rx $ and $1.05\rx$ the coupling squared $|\gamma_2|^2$ changes by  16\% for $\rx=0.52$  and by  40\% for $\rx=0.75$. This implies that though there is no a critical dependence of $|\gamma_2|$ on the value of $\rx$ as in the case of the $f_0(980)$ from Ref.~\cite{Guo:2012yt}, cf. Table~\ref{tabflatt2}, it is also true that the variations are important. Considering the experimental errors in $\rx$ from Refs.~\cite{Aubert:2006nu,BES:2005iaq} of around a 23\% for $\rx=0.52$ and $15\%$ for $\rx=0.75$,  we have added in quadrature the resulting uncertainty of around a 40\%   in $X_2$ to the one stemming from the values of $M_R$ and $\Gamma_R$ for calculating the error bars in Table~\ref{tab.211111.1}.

A similar relative uncertainty, which also follows by inspection of the dashed line in the left panel of Fig.~\ref{pic_X_r},  has been applied to $X_2$ in Table~\ref{tabff3} calculated with the method S. We have  proceeded  analogously for the error bars of $X_2$ shown in Table~\ref{tab.211106.2}. Notice that within the method S the sensitivity to the input value of $\rx$ for the pole in Eq.~\eqref{211107.1} is a factor of 1.6 larger than for the pole in Eq.~\eqref{equ22}, as follows by comparing the slope of the solid versus the dashed lines in the left panel of Fig.~\ref{pic_X_r}.

We then find a rather unpleasant situation in which small changes in the input values, perfectly well within the error bars provided by the analyses where they come from, can give rise to a very different output value for $|\gamma_2|^2$ and then for $X_2$, the partial compositeness coefficient which typically almost saturates the whole $X$.
A way that we have found to circumvent this limitation is to use the spectral density function $\omega(E)$, Eq.~\eqref{211013.1}, in order to calculate the compositeness $X$ as $1-W_R$, since we have checked that it does not depend so sensitively on the input value of $\rx$. In this regard, we calculate $W_R$ as a function of the extent of the interval $[-\Delta,\Delta]$ used in Eq.~\eqref{211013.1} because its dependence on $\Delta$ is worth noticing, and the results for the $f_0(980)$ pole in Eq.~\eqref{211107.1} are given in Table~\ref{tabflatt5g}. In the first column we show the $r_{\rm exp}$ taken, the integration interval in Eq.~\eqref{211013.1} is given in the second column, the third column is for the resulting $W_R$, the fourth column provides $1-W_{R}$, and the last one gives $X$ calculated already with the method based on the Flatt\'e parametrization in Table~\ref{tabflatt2}.

\begin{center}
\begin{figure}
\centering
\caption{ The spectral density function $\omega(E)$ is shown for the $f_0(980)$ in the top panels (a) and (b), and for the $a_0(980)$ in the bottom ones (c) and (d). For the $f_0(980)$ the poles considered are in the RS II, with the pole in Eq.~\eqref{211107.1} corresponding to the panel (a) and the pole in Eq.~\eqref{equ22} to the panel (b). For the $a_0(980)$ we take the poles in Eqs.~\eqref{211009.1} and ~\eqref{eq:ratioa0}, which are in the RS II (panel (c)) and RS III (panel (d)), respectively. Every line in a plot corresponds to the indicated value of $r_{\rm exp}$ in the legends, and for the $a_0(980)$ the poles lie in different RSs.
\label{picflatt}}
\begin{minipage}[t]{1\linewidth}
\centering
\subfigure[]{
\includegraphics[width=0.4\textwidth]{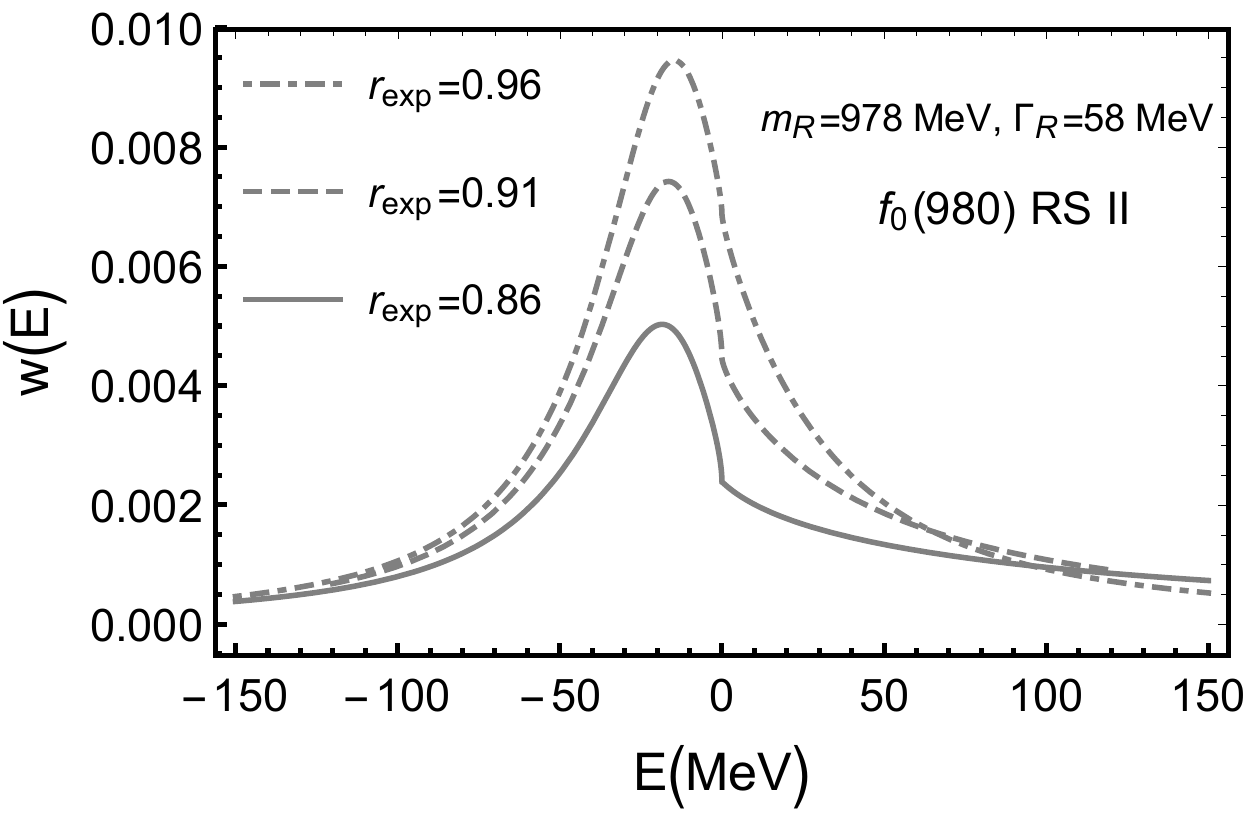}\;}
\hspace{1in}
\subfigure[]{
\includegraphics[width=0.4\textwidth]{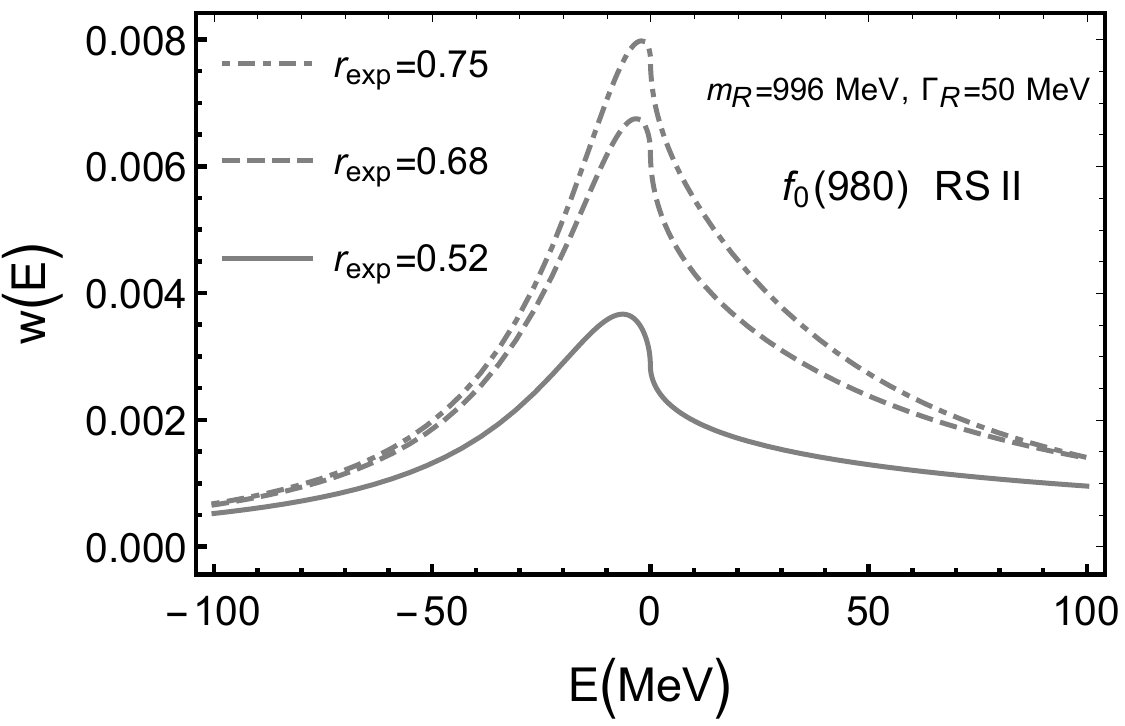}}
\subfigure[]{
\includegraphics[width=0.4\textwidth]{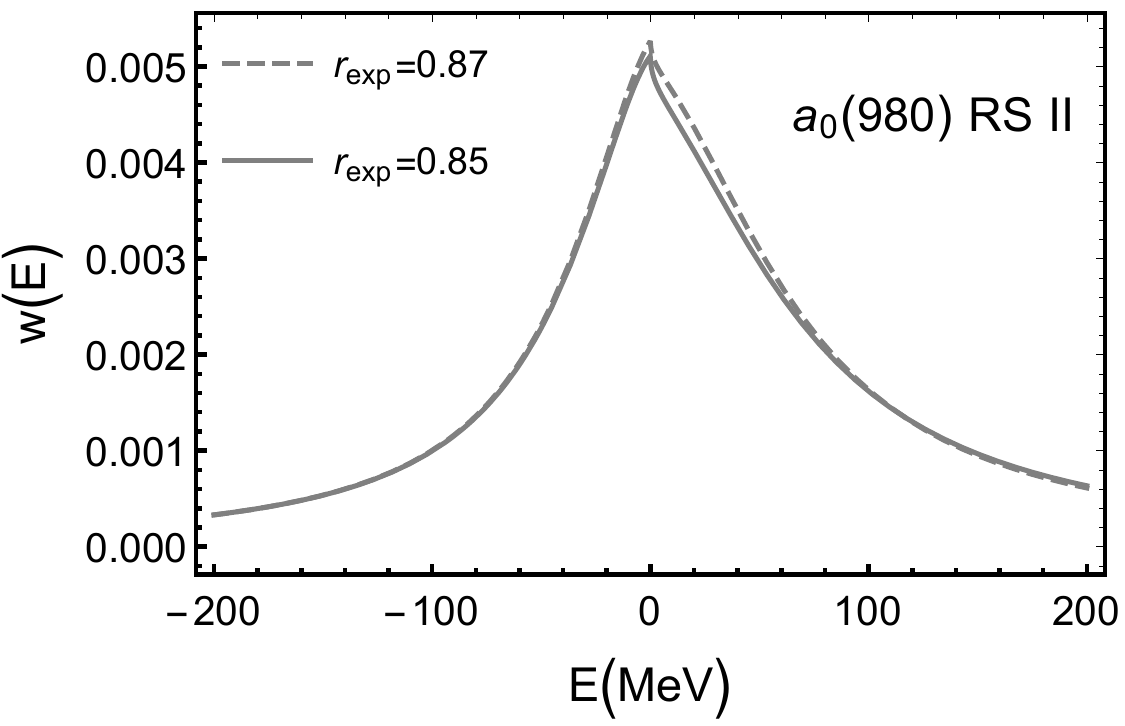}\;}
\hspace{1in}
\subfigure[]{
\includegraphics[width=0.4\textwidth]{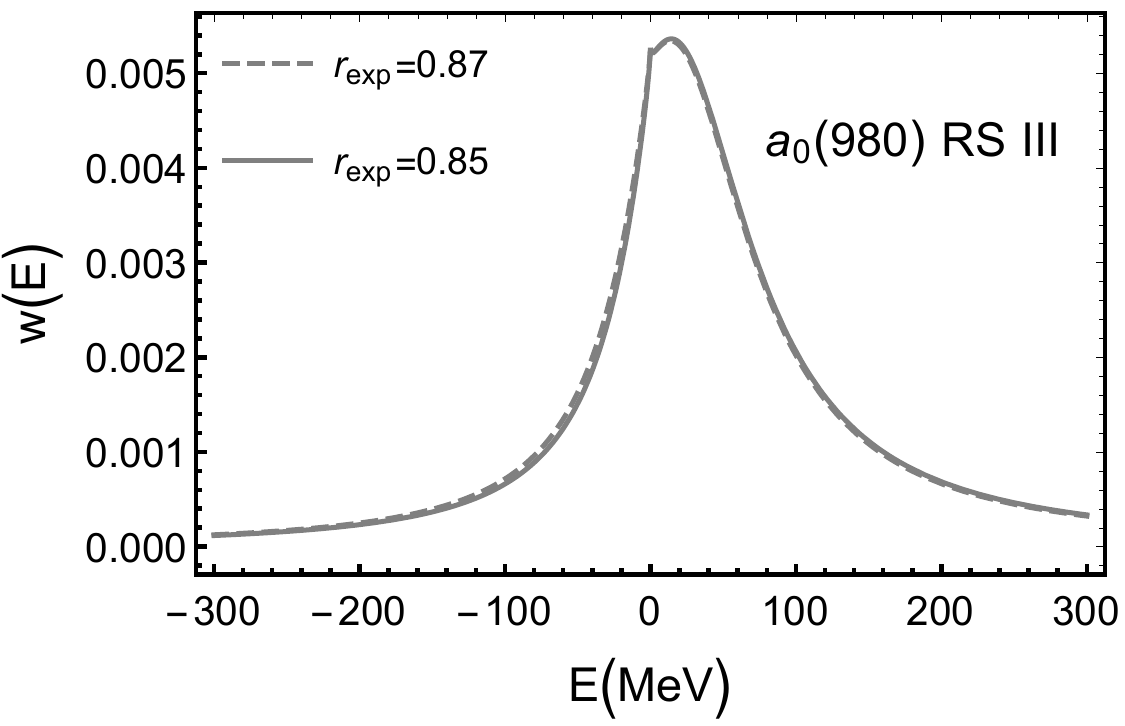}}
\end{minipage}
\end{figure}
\end{center}

In Ref.~\cite{Baru:2003qq} the cutoff $\Delta$ was taken to be 50~MeV. Since the width of the $f_0(980)$ in Ref.~\cite{Guo:2012yt} is 58~MeV we take $\Delta=60$~MeV for this pole, but estimate an uncertainty for the result of $W_R$ by considering also $\Delta=90$~MeV, which is a 50\% larger than the nominal value.
This is illustrated in Fig.~\ref{picflatt}, where we plot  $\omega(E)$ for the $f_0(980)$ and $a_0(980)$ resonances in the top and bottom panels, respectively. It is clearly seen that most of the resonance bump lies in the region $|E|<\Gamma_R$, and that for $|E|>1.5\Gamma_R$ it has already faded away. From Table~\ref{tabflatt5g} we obtain that for $\Delta=60$~MeV and $\rx=0.91$ (the nominal one for \cite{Guo:2012yt}) the resulting compositeness is $1-W_{f_0}=0.52$, and for $\Delta=1.5\Gamma_R\approx 90$~MeV, it decreases up to 0.43. We observe a variation of around a 40\% for $1-W_{f_0}$ in Table~\ref{tabflatt5g} calculated with $\Delta=60$~MeV (much smaller than the 300\% in Table~\ref{tabflatt2}), decreasing from 0.68 for $\rx=0.86$  to 0.39 for $\rx=0.96$.

\begin{table}[!htbp]
\begin{center}
\caption{ Resonance $f_0(980)$ with the pole position of  Eq.~\eqref{211107.1} in the RS \Rmnum{2}: We show the dependence of $W_{f_0}$ on the integration interval $[-\Delta,\Delta]$ with $\Delta$ up to $2\Gamma_R$. The last column gives $X=X_1+X_2$ obtained in Table~\ref{tabflatt2}. The values of $\rx$ given do not come from any experiment and are considered due to theoretical reasons. }
\label{tabflatt5g}
\begin{tabular}{|c|ccc|c|}
\Xhline{1pt}
~~$r_{\rm exp}$&~~~$[-\Delta, \Delta]$&~~~$W_{f_{0}}$ & $1-W_{f_0}$ & $X$ \\
\Xhline{1pt}
\multirow{4}*{0.86}~~~&~~~$[-30, 30 ]$~~~&~~~$0.19$~~~& 0.81 & \\
&~~~$[-60, 60 ]$~~~&~~~$0.32$~~~& 0.68 & \\
&~~~$[-90, 90 ]$~~~&~~~$0.40$~~~& 0.60 &  \\
&~~~$[-120, 120 ]$~~~&~~~$0.45$~~~& 0.55 & $0.36\pm0.14$\\
\Xhline{1pt}\multirow{4}*{0.91}~~~&~~~$[-30, 30 ]$~~~&~~~$0.30$~~~& 0.70 & \\
&~~~$[-60, 60 ]$~~~&~~~$0.48$~~~& 0.52 & \\
&~~~$[-90, 90 ]$~~~&~~~$0.57$~~~& 0.43 &  \\
&~~~$[-120, 120 ]$~~~&~~~$0.63$~~~& 0.37 & $0.27\pm0.12$\\
\Xhline{1pt}
\multirow{4}*{0.96}~~~&~~~$[-30, 30 ]$~~~&~~~$0.40$~~~& 0.60 & \\
&~~~$[-60, 60 ]$~~~&~~~$0.61$~~~& 0.39 & \\
&~~~$[-90, 90 ]$~~~&~~~$0.71$~~~& 0.29 &  \\
&~~~$[-120, 120 ]$~~~&~~~$0.76$~~~& 0.24 & $0.13\pm0.06$\\

\Xhline{1pt}
\end{tabular}
\end{center}
\end{table}

\begin{table}[!htbp]
\begin{center}
\caption{{\small Resonance $f_0(980)$ with the pole position in the RS \Rmnum{2} from  Ref.~\cite{GarciaMartin:2011jx}, Eq.~\eqref{equ22}: We show the dependence of $W_{f_0}$ on the integration interval $[-\Delta,\Delta]$ with $\Delta$ up to $2\Gamma_R$.  For the experimental inputs of $\rx$ we give our final estimate for $1-W_{f_0}$ in the column 5. The total compositeness $X=X_1+X_2$ from Table~\ref{tab.211111.1} is given in the last column. }
  \label{tabflatt5}}
\begin{tabular}{|c|ccc|c|c|}
\Xhline{1pt}
~~$r_{\rm exp}$&~~~$[-\Delta, \Delta]$&~~~$W_{f_{0}}$ & $(1-W_{f_0})_\Delta$  & $1-W_{f_0}$& $X$ \\

\Xhline{1pt}
\multirow{4}*{0.52
\cite{Aubert:2006nu}}~~~&~~~$[-25, 25 ]$~~~&~~~$0.13$~~~& 0.87 & &\\
&~~~$[-50, 50 ]$~~~&~~~$0.21$~~~& $0.79$ & &\\
&~~~$[-75, 75 ]$~~~&~~~$0.27$~~~& 0.73 &  &\\
&~~~$[-100, 100 ]$~~~&~~~$0.31$~~~& 0.69 & $0.76\pm 0.15$ &$0.51\pm0.22$\\
\Xhline{1pt}\multirow{4}*{0.68
\cite{Oller:1997ti}}~~~&~~~$[-25, 25 ]$~~~&~~~$0.25$~~~& 0.75 & &\\
&~~~$[-50, 50 ]$~~~&~~~$0.39$~~~& 0.61 & &\\
&~~~$[-75, 75 ]$~~~&~~~$0.47$~~~& 0.53 & & \\
&~~~$[-100, 100 ]$~~~&~~~$0.53$~~~& 0.47 & & $0.38\pm0.18$\\
\Xhline{1pt}
\multirow{4}*{0.75
\cite{BES:2005iaq}}~~~&~~~$[-25, 25 ]$~~~&~~~$0.30$~~~& 0.70 & & \\
&~~~$[-50, 50 ]$~~~&~~~$0.45$~~~& $0.55$ & &\\
&~~~$[-75, 75 ]$~~~&~~~$0.55$~~~& 0.45 & & \\
&~~~$[-100, 100 ]$~~~&~~~$0.61$~~~& 0.39 & $0.50\pm 0.15$ & $0.29\pm0.14$\\
\Xhline{1pt}
\end{tabular}
\end{center}
\end{table}

Next, we apply the method based on the spectral density function $\omega(E)$ to the pole from Ref.~\cite{GarciaMartin:2011jx}, Eq.~\eqref{equ22}, and  the results are given in  Table~\ref{tabflatt5}. The dependence with $\Delta$ in the compositeness $1-W_{f_0}$ is indicated with the subscript $\Delta$ in the fourth column of Table~\ref{tabflatt5}, $(1-W_{f_0})_\Delta$.
We take as  the nominal value for $1-W_{f_0}$ the mean between the ones obtained with $\Delta=\Gamma_R=50$~MeV and $\Delta=1.5\Gamma_R$.
Considering the variation of the results between these two values for $\Delta$, and the error bar of around 0.12 in $\rx$ in both experiments \cite{Aubert:2006nu,BES:2005iaq} (such that $\rx=0.68$ is around the upper value of $\rx=0.52$ and the lower one for 0.75 within one standard deviation), we see that an uncertainty of at least 0.15 should be considered for {the reference value} of $1-W_{f_0}$.
The final figure is given in the column before the last one in Table~\ref{tabflatt5}.
We can appreciate that the central value obtained by integrating Eq.~\eqref{211013.1} is larger than the central value for $X$ in Table~\ref{tab.211111.1} by employing Eq.~\eqref{equ1},  which is also given in the last column.
 Then, we would have $1-W_{f_0}=0.76\pm 0.15$ and $0.50\pm 0.15$ for $\rx=0.52$ \cite{Aubert:2006nu} and $0.75$ \cite{BES:2005iaq}, respectively, as also shown in Table~\ref{tabflatt5}. These figures indicate a dominant meson-meson component, mostly $K\bar{K}$ ($X_2\gg X_1$), in the nature of the $f_0(980)$ pole  from Ref.~\cite{GarciaMartin:2011jx}, Eq.~\eqref{equ22}. It also follows that the results are  compatible with those given for $X$ within errors.

However, for $\rx$ from Ref.~\cite{BES:2005iaq} the central value of $1-W_{f_0}$ in Table~\ref{tabflatt5} is only slightly above 0.5, and both  values taken for $\rx$ \cite{Aubert:2006nu,BES:2005iaq} generate values of $1-W_{f_0}$ that could decrease substantially once errors are taken into account. Therefore, other  components apart from  $K\bar{K}$ are likely to play a noticeable role in the composition of the $f_0(980)$.
From nonperturbative studies based on unitarizing chiral perturbation theory with/without resonances these extra components have been also unveiled \cite{Oller:1998zr,Pelaez:2006nj,Guo:2012ym,Guo:2012yt}.

\subsection{The $a_0(980)$ resonance}
\label{sec.211126.2}

Now, we follow similar steps and use  the fact that the ratio of $\Gamma(a_0(980)\to K\bar{K})/\Gamma(a_0(980)\to \pi\eta)$ is also given by  Ref.~\cite{CrystalBarrel:2019zqh}, see Eqs.~\eqref{211009.1} and \eqref{eq:ratioa0}, with the values $0.138\pm 0.035$ (RS \Rmnum{2}) and $0.149\pm 0.039$  (RS \Rmnum{3}), respectively. The typical uncertainty for the inferred $\rx$ is around a 3\%.  We also consider the average value given by the PDG \cite{Zyla:2020zbs}, $\Gamma(a_0(980)\to K\bar{K})/\Gamma(a_0(980)\to \pi\eta)=0.177\pm 0.024$, which implies an uncertainty of a 1.7\% in $\rx$.

Its implication has been organized in Table \ref{tabff5}, where we apply the method S with $\rx$ as input to the $a_0(980)$ poles in the RS \Rmnum{2} and \Rmnum{3}.
The $K\bar{K}$ component is larger than the $\pi\eta$ one,
but the total compositeness $X$ is small being less than 0.25 and 0.35 for the RS \Rmnum{2} and \Rmnum{3} $a_0(980)$ poles, respectively.

\begin{table}[!htbp]
\begin{center}
\caption{ Method S applied to the resonance $a_0(980)$ with pole positions from Ref.~\cite{CrystalBarrel:2019zqh},  Eqs.~\eqref{211009.1} and \eqref{eq:ratioa0}: By reproducing the input values of  $\Gamma_{a_0}$ and  $r_{\rm exp}$ (column 2), we predict $|\ga_i|$ (columns 3, 4),   $\Gamma_i$ (columns 5, 6), and  $X_i$ (columns 7, 8).}
\begin{tabular}{|cc|cccccc|}
\Xhline{1pt}
~~$r_{\rm exp}$&~~~$\text{RS}~~~$&~~~$|\gamma_{\pi\eta}|(\text{GeV})$~~~&$|\gamma_{K\bar{K}}|(\text{GeV})$~~~&$\Gamma_{1}(\text{MeV})$&~~~$\Gamma_{2}(\text{MeV})$&~~~$X_{\pi\eta}$&~~~$X_{K\bar{K}}$~~~\\
\Xhline{1pt}
\multirow{2}*{0.85 \cite{Zyla:2020zbs}}&~~~$\text{\Rmnum{2}}~~~$&~~~$2.90\pm0.05$~~~&$2.32\pm0.08$~~~&$111.8\pm4.2$&~~~$14.6\pm0.6$&~~~$0.100\pm0.004$&~~~$0.132\pm0.013$~~~\\
&~~~$\text{\Rmnum{3}}~~~$&~~~$2.86\pm0.05$~~~&$2.61\pm0.72$~~~&$108.0\pm3.7$&~~~$19.1\pm0.7$&~~~$0.098\pm0.004$&~~~$0.249\pm0.017$~~~\\
\Xhline{1pt}
\multirow{2}*{0.87 \cite{CrystalBarrel:2019zqh}}&~~~$\text{\Rmnum{2}}~~~$&~~~$2.88\pm0.05$~~~&$2.16\pm0.07$~~~&$109.8\pm4.2$&~~~$12.6\pm0.5$&~~~$0.098\pm0.004$&~$0.115 \pm 0.009$ \\
&~~~$\text{\Rmnum{3}}~~~$&~~~$2.89\pm0.05$~~~&$2.43\pm0.07$~~~&$110.5\pm3.8$&~~~$16.5\pm0.6$&~~~$0.100\pm0.004$&~~~~~$0.216\pm 0.030$ ~~~ \\ 
\Xhline{1pt}
\end{tabular}
\label{tabff5}
\end{center}
\end{table}

\begin{table}[!htbp]
\begin{center}
\caption{ Method F applied to the resonance $a_0(980)$ with pole positions from Ref.~\cite{CrystalBarrel:2019zqh}, Eqs.~\eqref{211009.1} and \eqref{eq:ratioa0}: The branching ratio $r_{\text{exp}}$ is taken as input (column 1), the RS in which the pole lies is indicated in the column 2. We then calculate the bare partial-decay width $\widetilde{\Gamma}_1$ (column 3), the bare coupling squared $g_2$ (column 4), $E_f$ (column 5),  the renormalized $\Gamma_1$ (column 6) and $\Gamma_2$ (column 7), and the partial compositeness coefficients $X_1$ (column 8) and  $X_2$ (column 9).  }
\begin{tabular*}{\linewidth}{@{\extracolsep{\fill}}|cc|ccccccc|}  
\Xhline{1pt}
$r_{\text{exp}}$ &$\text{RS}$ &$\widetilde{\Gamma}_{\pi\eta}\,(\text{MeV})$ &$g_2$ &$E_{f}\,(\text{MeV})$ &$\Gamma_{1}\,(\text{MeV})$ &$\Gamma_{2}\,(\text{MeV})$ &$X{\pi\eta}$ &$X_{K\bar{K}}$\\
\Xhline{1pt}
\multirow{2}*{0.85 \cite{Zyla:2020zbs}} &$\text{\Rmnum{2}}$ &$124.0\pm5.3$ &$0.21\pm0.03$ &$3.08\pm5.11$& $111.8\pm4.2$ &$14.6\pm0.6$ &$0.100\pm0.004$ &$0.116\pm 0.017$ \\ 
&$\text{\Rmnum{3}}$ &$98.7\pm3.4$ &$0.21\pm0.01$ &$23.1\pm2.8$ &$108.0\pm3.6$ &$19.1\pm0.6$ &$0.099\pm0.004$ &$0.205\pm 0.028$ \\ 
\Xhline{1pt}
\multirow{2}*{0.87 \cite{CrystalBarrel:2019zqh}} &$\text{\Rmnum{2}}$ &$119.9\pm4.9$ &$0.18\pm0.02$ &$4.65\pm4.82$ &$109.8\pm4.2$ &$12.6\pm0.5$ &$0.098\pm0.004$ & $0.100\pm 0.016$ \\ 
&$\text{\Rmnum{3}}$ &$102.0\pm3.5$ &$0.18\pm0.01$ &$21.6\pm2.9$ &$110.5\pm3.7$ &$16.5\pm0.6$ &$0.101\pm0.004$ &$0.178\pm 0.038$ \\
\Xhline{1pt}
\end{tabular*}
\label{tabflatt4}
\end{center}
\end{table}

We now  proceed with the application of the  method F with $\rx$ as input and the results obtained are presented in  Table~\ref{tabflatt4}. We observe that the values of $X_2$ and $\Gamma_{K\bar{K}}$ are rather similar between Tables~\ref{tabflatt3} and \ref{tabflatt4}, and similarly if we compare the latter with Table~\ref{tabff5} where $\rx$ is taken as input too. Thus, we  find compatibility between different methods for the $a_0(980)$ case.
It is also clear from Table~\ref{tabflatt4} that the value of $\rx=0.85$ obtained from the PDG average value of $\Gamma(a_0(980)\to K\bar{K})/\Gamma(a_0(980)\to \pi\eta)=0.177\pm 0.024$ \cite{Zyla:2020zbs}  implies quite small values for $X_2\lesssim 0.3$.

It seems in our opinion  that  an uncertainty of only $2\%-3\%$ in $\rx$ for the $a_0(980)$ is probably too optimistic. In this respect we notice that the analysis of Ref.~\cite{CrystalBarrel:2019zqh} for the $a_0(980)$ is based on taking poles in the RS II and III for this resonance, while  recent sophisticated theoretical studies \cite{Guo:2016zep,Dudek:2016cru}, which also reproduce lattice QCD data, require a very different qualitative picture with a pole in the RS IV.
In this respect, let us check the sensitivity of the results based on Eq.~\eqref{equ1} for the calculation of $X$, and for that take e.g.  the $a_0(980)$ pole in the RS \Rmnum{2} and  $\rx=0.85\pm 0.15$, which corresponds to the average  value of the PDG with an {\it ad hoc} uncertainty of around a 20\%.
We indeed find a strong sensitivity, such that  for the central value $\rx=0.85$ we have the values for the couplings $|\gamma_1|=2.9$~GeV and $|\gamma_2|=2.2$~GeV, while for the lower end $\rx=0.85-0.15=0.70$ we find a new solution with the values $|\gamma_1|=3.1$~GeV and $|\gamma_2|=3.2$~GeV.
With respect to the central value we have a variation of only a 6\% in $|\gamma_1|$, but $|\gamma_2|$ is now a 44\% bigger (a factor of 2 for the square of the coupling).
The variation is of similar size if considering $\rx=0.87\pm 0.17$ with the central value from Ref.~\cite{CrystalBarrel:2019zqh} and  an {\it ad hoc} $20\%$ uncertainty taken. 

Thus, it is advisable to also  apply for the
$a_0(980)$ case the method based on the spectral density function (which is less sensitive to small variations in the input value for $\rx$) and evaluate the compositeness $1-W_{a_0}$  as a  function of $\Delta$. We give the results in Tables~\ref{tab.211111.2} and \ref{tabflatt6} for the $a_0(980)$ poles in the RS \Rmnum{2}, Eq.~\eqref{211009.1},  and RS \Rmnum{3}, Eq.~\eqref{eq:ratioa0}, respectively. The dependence with $\Delta$ is indicated by the subscript $\Delta$ in the fourth column, $(1-W_{a_0})_\Delta$.
Taking into account the variation in the value of $1-W_{f_0}$ between $\Delta\approx \Gamma_{a_0}$ and  $\Delta\approx 1.5 \Gamma_{a_0}$, we give our range of values calculated for $1-W_{a_0}$ in the column before the last one in Tables~\ref{tab.211111.2} and \ref{tabflatt6}. The output is similar in both tables with values for $1-W_{a_0}$ typically within the range $0.3-0.4$. When compared with $X$ from Table~\ref{tabflatt4}, given in the last column in Tables~\ref{tab.211111.2} and \ref{tabflatt6}, we see a quantitative agreement in the case of the RS \Rmnum{3} pole, and a semiquantitative one for the RS \Rmnum{2} one. The emerging  picture is that $X$ is clearly less than 0.5, ranging between $0.2-0.4$ depending on the method of calculation. Therefore, other components beyond $\pi\eta$ and $K\bar{K}$ are also required  \cite{Dai:2012kf,Sekihara:2014qxa}. However, if the resonance lied in RS \Rmnum{4}, as preferred by the recent analyses \cite{Guo:2016zep,Dudek:2016cru}, our approach does not apply and we cannot extend such conclusion to that case.

\begin{table}[H]
\begin{center}
  \caption{Resonance $a_0(980)$ with the pole position in the RS \Rmnum{2} from Ref.~\cite{CrystalBarrel:2019zqh},  Eq.~\eqref{211009.1}. The dependence of $W_{a_0}$ on the integration interval $[-\Delta,\Delta]$ for the $a_{0}(980)$ is shown with  $\Delta$ up to $2\Gamma_{a_0}$. In the column 5 we provide our interval estimated for $1-W_{a_0}$ and in the last one $X=X_1+X_2$ from Table~\ref{tabflatt4} is given.
    \label{tab.211111.2}}
\begin{tabular}{|c|ccc|c|c|}
\Xhline{1pt}
~~$r_{\rm exp}$&~~~$[-\Delta, \Delta]$&~~~$W_{a_{0}}$& $(1-W_{a_0})_{\Delta}$ & $1-W_{a_0}$ & $X$\\

\Xhline{1pt}
\multirow{4}*{~~~0.85 \cite{Zyla:2020zbs}}~~~&~~~$[-50, 50 ]$~~~&~~~~$0.38$~~~ & 0.62 & &\\
~~~&~~~$[-100, 100 ]$~~~&~~~~$0.57$~~~ & $0.43$ & &\\
~~~&~~~$[-150, 150 ]$~~~&~~~$0.67$~~~& 0.33 & &\\
~~~&~~~$[-200, 200 ]$~~~&~~~$0.73$~~~& 0.27 & $0.33-0.43$ &$0.216 \pm 0.017$\\
\Xhline{1pt}

\multirow{4}*{~~~0.87 \cite{CrystalBarrel:2019zqh}}~~~&~~~$[-50, 50 ]$~~~&~~~~$0.39$~~~ & 0.61 & &\\
~~~&~~~$[-100, 100 ]$~~~&~~~$0.59$~~~& $0.41$ & &\\
~~~&~~~$[-150, 150 ]$~~~&~~~$0.68$~~~& 0.32 & &\\
~~~&~~~$[-200, 200 ]$~~~&~~~$0.74$~~~& 0.26 & $0.32-0.41$ &$0.198 \pm 0.016$\\
\Xhline{1pt}
\end{tabular}
\end{center}
\end{table}

\begin{table}[H]
\begin{center}
  \caption{ Resonance $a_0(980)$ with the pole position in the RS \Rmnum{3} from Ref.~\cite{CrystalBarrel:2019zqh},  Eq.~\eqref{eq:ratioa0}. The dependence of $W_{a_0}$ on the integration interval $[-\Delta,\Delta]$ for the $a_{0}(980)$ is shown with  $\Delta$ up to $2\Gamma_{a_0}$. In the column 5 we provide our interval estimated for $1-W_{a_0}$ and in the last one $X=X_1+X_2$ from Table~\ref{tabflatt4} is given.
  \label{tabflatt6}}
\begin{tabular}{|c|ccc|c|c|}
\Xhline{1pt}
~~$r_{\rm exp}$&~~~$[-\Delta, \Delta]$&~~~$W_{a_{0}}$& $(1-W_{a_0})_\Delta$ & $1-W_{a_0}$ &  $X$\\

\Xhline{1pt}
\multirow{5}*{~~~0.85 \cite{Zyla:2020zbs}}~~~&~~~$[-50, 50 ]$~~~&~~~~$0.39$~~~ & 0.61 & & \\
~~~&~~~$[-100, 100 ]$~~~&~~~~$0.59$~~~ & $0.41$ & & \\
~~~&~~~$[-150, 150 ]$~~~&~~~$0.69$~~~& 0.31 & &\\
~~~&~~~$[-200, 200 ]$~~~&~~~$0.75$~~~& 0.25 & &\\
~~~&~~~$[-250, 250 ]$~~~&~~~$0.79$~~~& 0.21 & $0.31-0.41$ &$0.303\pm0.030$\\
\Xhline{1pt}

\multirow{5}*{~~~0.87 \cite{CrystalBarrel:2019zqh}}~~~&~~~$[-50, 50 ]$~~~&~~~~$0.40$~~~ & 0.60 & &\\
~~~&~~~$[-100, 100 ]$~~~&~~~$0.60$~~~& $0.40$ & &\\
~~~&~~~$[-150, 150 ]$~~~&~~~$0.70$~~~& 0.30 & &\\
~~~&~~~$[-200, 200 ]$~~~&~~~$0.76$~~~& 0.24 & &\\
~~~&~~~$[-250, 250 ]$~~~&~~~$0.80$~~~& 0.20 & $0.30-0.40$ &$0.279\pm 0.037$\\ 
\Xhline{1pt}
\end{tabular}
\end{center}
\end{table}

We would also like to comment about  the clearly visible cusp effect for most of the curves of $\omega(E)$ in Fig.~\ref{picflatt}. This change in the shape of $\omega(E)$ below and above the two-kaon threshold is due to the fact that if the resonance pole lies in the RS \Rmnum{2} (\Rmnum{3}) then  there is no associated pole in the RS \Rmnum{3} (\Rmnum{2}) above (below) the $K\bar{K}$ threshold. Precisely the RS III (II) is the one that connects with the physical region there.

\section{Summary and conclusions}
\label{sec.211106.1}

This paper discusses the importance of the continuum channels $\pi\pi$-$K\bar{K}$ and $\pi\eta$-$K\bar{K}$ in the composition of the $f_0(980)$ and $a_0(980)$ resonances, which is quantified by the concept of  the total compositeness $X$.
In our calculation we exploit the tight relationship between the compositeness $X$, the mass and the  decay width of a resonance.
The threshold of the $K\bar{K}$ pair is very close to the mass of each resonance and this fact has main consequences in our results. We develop two methods: One is based on saturating the total width and compositeness;   the other relies on the use of a Flatt\'e parametrization and, in some instances, of the spectral function of a near-threshold resonance.

We provide input values for the mass and width of each resonance by taking their pole positions from relevant analyses in the literature.
In particular, for the $f_0(980)$ we consider the determination of its pole position  by the Roy-like GKPY equations. 
Regarding the third input needed in our analyses we first take input values for $X$ in the compositeness relationship and
we predict the couplings, partial compositeness coefficients and partial-decay widths to the  $\pi\pi$ ($\pi\eta$) and $K\bar{K}$ channels for the $f_0(980)$ ($a_0(980))$.
There is an interesting trend in the results such that the larger $X$, the smaller the branching decay ratio to the lighter channel, $r_{\rm exp}$.
 This is due to the increase of the coupling to the heavier $K\bar{K}$ channel with increasing $X$, compensating the reduced phase space available for the decay of the resonances into this channel. It is also found that for the $f_0(980)$ the partial compositeness coefficient of $K\bar{K}$, $X_2$, is larger by orders of magnitude than the corresponding one to $\pi\pi$, $X_1$. For the $a_0(980)$ the compositeness for $K\bar{K}$ is also larger than the one associated to $\pi\eta$, but as $X$ decreases they tend to become similar in size.

Another possibility is to replace the third input $X$ by reported values in the literature for $r_{\rm exp}$.
However, if $X_2$ is calculated  in terms of the coupling squared  to $K\bar{K}$ and the derivative of the unitary-loop function in the corresponding Riemann sheet, we typically find a large sensitivity on the input value for $\rx$. The situation is improved when using the method based on integrating the spectral density function around the $K\bar{K}$ threshold along the  energy region comprising the resonance signal, so that more stable results are obtained under small changes in $\rx$.
It turns out that for the poles considered  the meson-meson components are typically dominant for the $f_0(980)$, while for the $a_0(980)$ they  are subdominant. 
By considering the range of nowadays acceptable values of $\rx$ for the $f_0(980)$ in the PDG \cite{Zyla:2020zbs} (within an interval of values of 0.4--0.9) the total compositeness $X$ could vary between 0.4 to 0.9 (accounting also for uncertainties in the values calculated).
It comes out unambiguously from our results that $X$ is  dominated by far by the $K\bar{K}$ component over the $\pi\pi$ one.
Regarding the $a_0(980)$ described by a pole in the RS II or III, as it is the case in Flatt\'e or Breit-Wigner like parametrizations, $\rx=0.85\pm 0.015$, which corresponds to the PDG average value for $\Gamma(a_0\to\pi\eta)/\Gamma(a_0\to K\bar{K})=0.177\pm 0.024 $, implies remarkably low values for the compositeness between 0.3 to 0.4 only.
In this case, we also find that the $K\bar{K}$ dominates over the $\pi\eta$ component but not overwhelmingly. 

Throughout the manuscript we have emphasized the need to distinguish  in a Flatt\'e parametrization between the bare  couplings/widths, on the one hand, and the dressed/renormalized ones, on the other hand. We have also shown how to calculate the latter ones.  In addition, we  discuss the relationship between the partial-decay widths directly calculated in terms of the dressed couplings and the actually  measured ones. In this regard, we show the changes needed for a pole in the second Riemann sheet lying near the heavier threshold, such that the total width is then $\Gamma_R=\Gamma_1- \Gamma_2$, instead of the standard $\Gamma_R=\Gamma_1+\Gamma_2$ for a pole in the third Riemann sheet.

Finally, we stress that the compositeness concept, as a quantitative examination of the inner structure of a resonance/molecule, is a relevant tool to promote a step forward in the understanding of the structure of a hadronic state.

\acknowledgments
We would like to thank useful discussions with Zhi-Hui Guo, J.~R. Pel\'aez and J.~Ruiz de Elvira. The author X.W.K. is supported by the National Natural Science Foundation of China (NSFC) under Project No.~11805012. J.A.O. acknowledges partial financial support by the MICINN AEI (Spain) Grant No. PID2019-106080GB-C22/AEI/10.13039/501100011033, and by the EU Horizon 2020 research and innovation programme, STRONG-2020 project, under grant agreement No.~824093.

\bibliographystyle{apsrev4-1}
\bibliography{ref}

\end{document}